\documentclass[10pt,twocolumn,letterpaper]{article}

\usepackage{wacv}              

\usepackage{graphicx}
\usepackage{amsmath}
\usepackage{amssymb}
\usepackage{booktabs}
\usepackage{siunitx}
\usepackage{nicefrac, xfrac}
\usepackage{array}
\usepackage{numprint}
\usepackage{placeins}
\usepackage{listings}
\usepackage{multicol}

\usepackage[pagebackref,breaklinks,colorlinks]{hyperref}
\usepackage{glossaries}
\usepackage{algorithm}
\usepackage{algpseudocodex}
\usepackage[titletoc,toc,page]{appendix}
\glsdisablehyper
\usepackage[accsupp]{axessibility} 

\newacronym{GAN}{GAN}{Generative Adversarial Net}
\newacronym{WSI}{WSI}{Whole Slide Image}
\newacronym{DDPM}{DDPM}{Denoising Diffusion Probabilistic Models}
\newacronym{SR}{SR}{Super-Resolution}
\newacronym{LR}{LR}{Low-Resolution}
\newacronym{HR}{HR}{High-Resolution}
\newacronym{IR}{IR}{Improved Recall}
\newacronym{IP}{IP}{Improved Precision}
\newacronym{GT}{GT}{ground-truth}
\newacronym{NFE}{NFE}{neural function evaluations}
\newacronym{iddpm}{iDDPM}{improved Denoising Diffusion Probabilistic Models}
\newacronym{VP}{VP}{variance preserving}
\newacronym{VE}{VE}{variance exploding}
\newacronym{DDNM}{DDNM}{Denoising Diffusion Null-Space Model}
\newacronym{DNN}{DNN}{Deep neural network}
\newacronym{LDM}{LDM}{latent diffusion model}
\newacronym{SDE}{SDE}{stochastic differential equation}
\newacronym{ODE}{ODE}{ordinary differential equation}
\DeclareSIUnit\px{px}

\usepackage[capitalize]{cleveref}
\crefname{section}{Sec.}{Secs.}
\Crefname{section}{Section}{Sections}
\Crefname{table}{Table}{Tables}
\crefname{table}{Tab.}{Tabs.}
\crefname{appsec}{Appendix}{Appendices}

\def\wacvPaperID{1205} 
\def\confName{WACV}
\def\confYear{2024}

\DeclareMathAlphabet{\pazocal}{OMS}{zplm}{m}{n}

\DeclareMathOperator*{\argmin}{arg\,min}
\newcommand{\unif}{\pazocal{U}}

\begin{document}

\title{Diffusion-based generation of Histopathological Whole Slide Images at a Gigapixel scale}

\author{Robert Harb\textsuperscript{1,2}, Thomas Pock\textsuperscript{1}, Heimo Müller\textsuperscript{2}\\
\textsuperscript{1}Institute of Computer Graphics and Vision, Graz University of Technology, Austria\\
\textsuperscript{2}Diagnostic and Research Institute of Pathology, Medical University of Graz, Austria\\
{\tt\small \{robert.harb, pock\}@icg.tugraz.at, heimo.mueller@medunigraz.at}
}
\maketitle

\begin{abstract}
    We present a novel diffusion-based approach to generate synthetic histopathological Whole Slide Images (WSIs)
at an unprecedented gigapixel scale. Synthetic WSIs have
many potential applications:
    They can augment training datasets to enhance the performance of many computational pathology applications.
They allow the creation of synthesized copies of datasets that can be shared without violating privacy regulations.
    Or they can facilitate learning representations of WSIs without requiring data annotations.
    Despite this variety of applications, no existing deep-learning-based method generates WSIs at their typically high resolutions. Mainly due to the high computational complexity. Therefore, we propose a novel coarse-to-fine sampling scheme to tackle image generation of high-resolution WSIs. In this scheme, we increase the resolution of an initial low-resolution image to a high-resolution WSI. Particularly, a diffusion model sequentially adds fine details to images and increases their resolution. In our experiments, we train our method with WSIs from the TCGA-BRCA dataset. Additionally to quantitative evaluations, we
also performed a user study with pathologists.
The study results suggest that our generated WSIs resemble
the structure of real WSIs.
\end{abstract}


\section{Introduction}
Histopathology is the study of diseases through the inspection of tissue samples. 
It plays a vital role in clinical practice by providing information for accurate diagnosis. 
Furthermore, it is also essential in medical research for studying disease processes and contributing to developing new therapeutic strategies.

Histopathological analysis is preceded by a few preparatory steps.
One first collects tissue samples, \eg via biopsies, excisions, or endoscopies.
Then, the samples are fixed, encased in paraffin, and thinly sliced.
The resulting tissue slices are then mounted on glass slides. 
Followed by staining, \eg using hematoxylin and eosin (H\&E), to enhance the visibility of cellular components and highlight specific tissue features. 
After staining, slides can be scanned, resulting in high-resolution images, so-called \glspl{WSI}. 
Notably, a typical \gls{WSI} has resolutions in the gigapixel range.

    \begin{figure}[t]
        \begin{center}
            \includegraphics[width=0.43\textwidth]{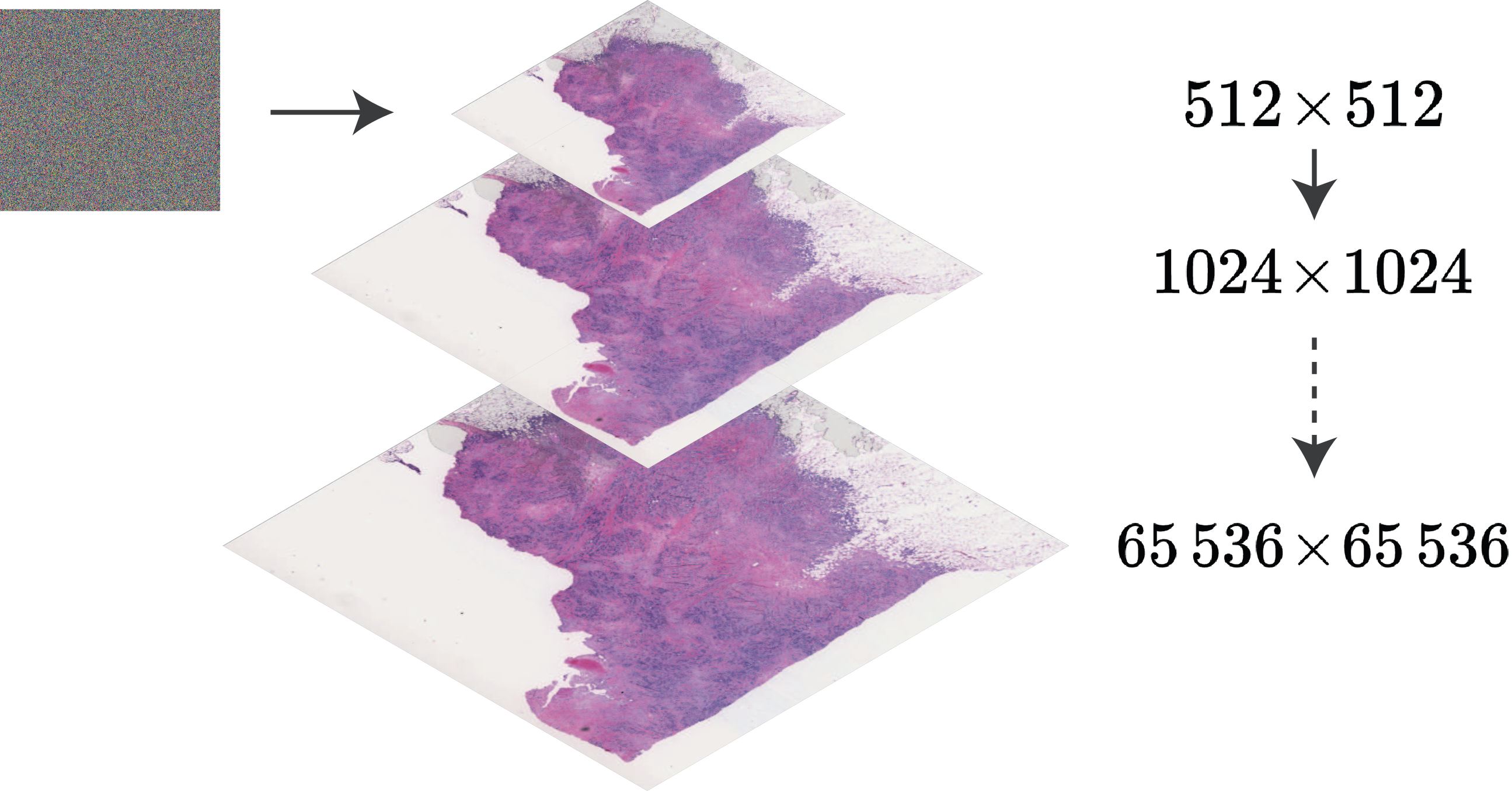}
        \end{center}
        \vspace{-1.5em}
        \caption{We sample a low-resolution image from noise using a diffusion-based generative image model.
        This low-resolution image is then sequentially upsampled in a coarse-to-fine scheme to generate a high-resolution Whole Slide Image.}
        \label{fig:coarsefine}
    \end{figure}

\FloatBarrier
\begin{figure*}[htb] 
    \centering
    \includegraphics[width=0.15\textwidth]{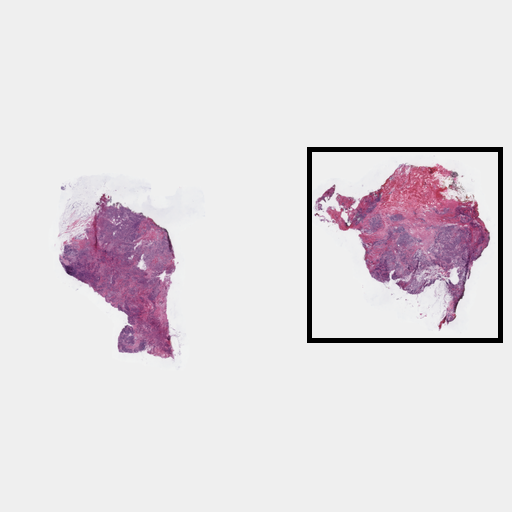}
    \hfill
    \includegraphics[width=0.15\textwidth]{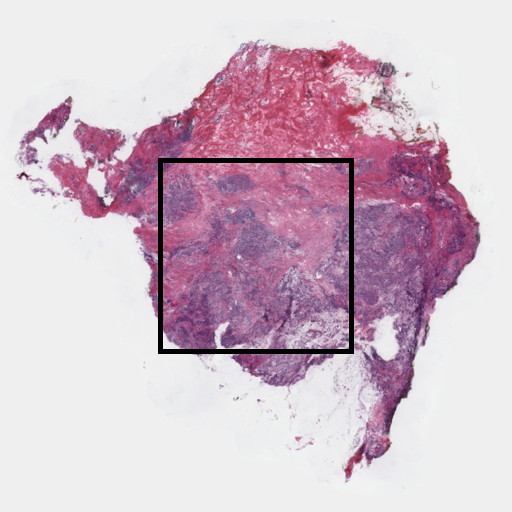}
    \hfill
    \includegraphics[width=0.15\textwidth]{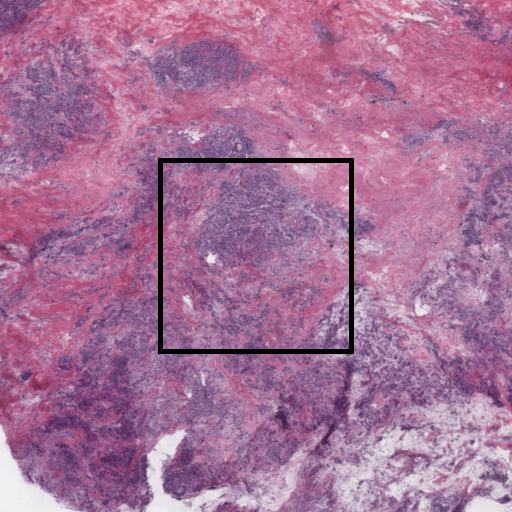}
    \hfill
    \includegraphics[width=0.15\textwidth]{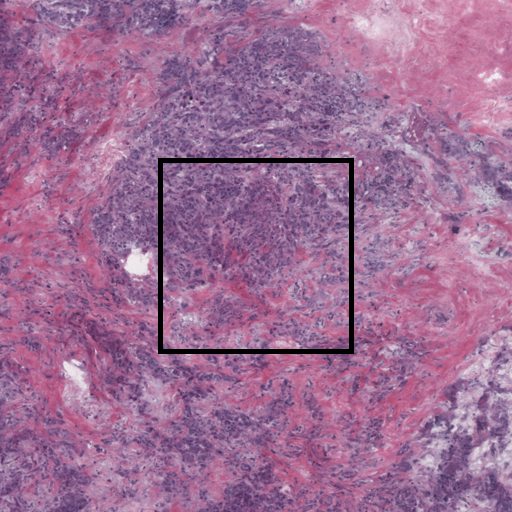}
    \hfill
    \includegraphics[width=0.15\textwidth]{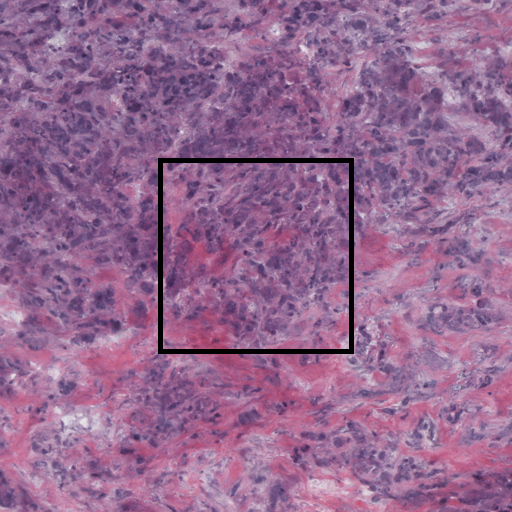}
    \hfill
    \includegraphics[width=0.15\textwidth]{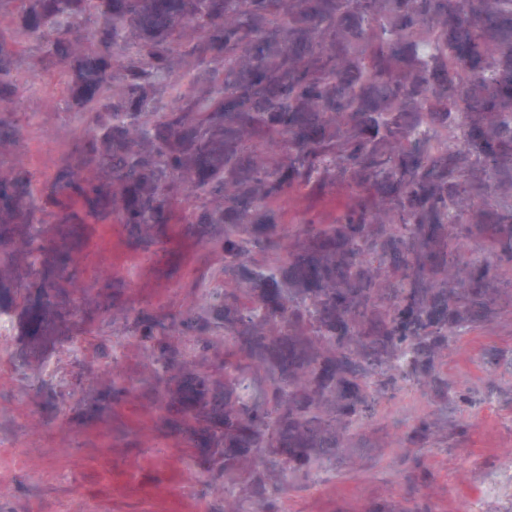}
    \caption{A synthetic \gls{WSI} generated by our method with a resolution of $\numprint{65536}\!\times\!\numprint{65361}$ pixels.
    Our synthesized WSI covers the whole magnification spectrum of a WSI, starting from a macroscopic overview of tissue, down to structures at the cellular level.
    For visualization, we extracted patches at different magnifications, the black rectangle shows the location of the patch in the subsequent column.
    }
    \label{fig:teaser}
\end{figure*}

A major challenge when developing algorithms that analyse \glspl{WSI} is their high resolution.
Many established methods are unsuitable since they are designed for much smaller resolutions.
This also applies in the field of synthetic image generation with deep-learning.
Although some methods exist, they all generated only small excerpts of \glspl{WSI}, \ie patches.
However, such low-resolution patches contain far less detail than entire \glspl{WSI}.
Their high resolution offers a spectrum of detail, from a macroscopic overview of the tissue sample to fine details like individual cells at the highest magnification.
Having this breadth of information is essential for many pathological applications.
Consequently, to fully harness the potential of synthetic data in histopathology, generating \glspl{WSI} at their full resolution is crucial.

There are many applications that could benefit from synthetic WSIs.
For instance, using synthetic data to augment datasets is common to improve the performance of deep-learning models, \eg, in segmentation or classification \cite{nikolenko2021synthetic}. 
Both of these tasks are essential in computational pathology.
For example, to find new Biomarkers \cite{wulczyn2021interpretable}, make survival predictions \cite{echle2021deep}, or for tumor segmentation \cite{vu2019methods}.

Moreover, synthetic \glspl{WSI} could unlock access to currently inaccessible datasets for broad audiences.
Despite institutions like biobanks or hospitals collect vast amounts of human tissue samples, data protection laws often prevent publishing this data without restrictions.
This limits the accessibility for research, hindering potential advancements in the field.
One approach to circumvent this issue is to publish synthesised versions of real datasets \cite{kaissis2020secure}.
Such synthesised datasets could maintain patient privacy while preserving the diagnostically relevant attributes of the original data.

Besides generating data, generative models can also be leveraged to learn data representations without requiring annotations \cite{radford2015unsupervised, hinton2006fast, wei2023diffusion}.
This is of particular interest in histopathology. Annotating WSIs is time-consuming due to their high resolutions and can often only be done by pathologists that have the necessary domain knowledge.

Motivated by the multitude of potential applications,
this work presents a novel diffusion-based method to generate synthetic \glspl{WSI}.
Most significantly, we generate \glspl{WSI} at remarkably high resolutions up to $\numprint{65536}\!\times\!\numprint{65536}$ pixels.
\cref{fig:teaser} shows such a high-resolution image generated by our approach.

The major challenge of our method is the computational infeasibility of training diffusion models for the high-resolution of \glspl{WSI}.
Instead, we are limited to a model that processes much lower-resolution images.
We tackle this limitation through a novel coarse-to-fine diffusion-based sampling scheme.
In this scheme, as illustrated in \cref{fig:coarsefine}, we sample a low-resolution image and step-wise increase its resolution.
Each step gradually adds finer details to an image while preserving its coarse structure.
While the initial image entirely fits into our model, we do the refinement patch-wise at later steps.
Even though patching limits the models' image context at later steps, the scheme has shown to be effective.
This is because the coarse image structure is established in the first steps, where the context is still large.
The refinement at later steps preserves this structure while gradually adding fine details that do not always require full-image context.

We describe our method in detail in \cref{sec:meth}.
The main contributions of our work are as follows:
\vspace{-0.5em}
\begin{itemize}
    \item To the best of our knowledge, we propose the first deep-learning-based method that creates synthetic histopathological WSIs at high resolutions up to $\numprint{65536}\!\times\!\numprint{65536}$ pixels.
    \vspace{-.5em}
    \item To this end, we propose a novel diffusion-based coarse-to-fine sampling scheme, where we guide the diffusion process with a relaxed super-resolution constraint.
    \vspace{-.5em}
    \item Even though our method involves patch-wise processing, we generate images without visible stitching artefacts.
    We achieve this through grid-shift, a novel technique where we interleave patching with diffusion iterations.
    In comparison with a related method, mask-shifting, grid-shift is computationally more efficient and simple to parallelize.
    \vspace{-.5em}
    \item We perform a user study with pathologists that suggests that our generated \glspl{WSI} are not consistently distinguishable from real \glspl{WSI}.
\end{itemize}

\label{sec:intro}

\section{Related Work}
In the following, we review related work in the areas of generating histopathological images and scaling diffusion models to high-resolutions.

\paragraph{Generation of Histopathology images.}
Several previously published methods tackle the generation of synthetic histopathological images.
However, our approach stands out as the only one that is able to generate \glspl{WSI} at gigapixel scale and is at the same time based on state-of-the-art generative deep-learning approaches.

A few methods were published before deep-learning-based image generation methods were widely adopted.
Instead, these methods \cite{synthesis_old, example-based} are based on texture-based image synthesis \cite{wei2009state, efros1999texture, heeger1995pyramid, portilla2000parametric}, where the synthesis process is based on the composition and modification of a small number of input patches.
However, this approach lacks generalizability and, instead of producing diverse content, mainly replicates the features of the few provided input patches.

Contrarily to texture-based image synthesis, deep learning-based image generation methods can learn complex patterns from large training datasets that allow them to generate diverse and realistic images.
This was demonstrated by several works \cite{levine2020synthesis, xue2021selective, dolezal2023deep} through the usage of \glspl{GAN} \cite{GANs}.
However, all of them only generated low-resolution patches and not high-resolution WSIs.

Though \glspl{GAN} have been the dominant approach to generate histopathological images, diffusion models are becoming increasingly popular in other domains \cite{croitoru2023diffusion}.
Mainly because \glspl{GAN} tend to be unstable at training \cite{lucic2018gans}, and suffer from mode collapse \cite{lin2018pacgan}.
Moreover, in many domains diffusion models have shown to outperform \glspl{GAN} \cite{dhariwal2021diffusion}, including medical images \cite{muller2022diffusion}.
Consequently, Moghadam \etal \cite{moghadam2023morphology} used diffusion for histopathology image generation.
However, in contrast to our work, only for small patches not for entire \glspl{WSI}.


\paragraph{Diffusion for high-resolution images.}
Training diffusion models \cite{pmlr-v37-sohl-dickstein15} is expensive, and the computational complexity grows with the image resolution.
Consequently, early works operated on low-resolution images up to $256\!\times\!256$ pixels \cite{ho2020denoising}.
Since then, various approaches have been proposed to enable generation of images with higher resolutions.
However, to the best of our knowledge, we are the first to demonstrate image generation with diffusion models at a gigapixel scale.

A common approach to scale diffusion models for higher resolutions are \glspl{LDM} \cite{rombach2022high}.
In \glspl{LDM}, the diffusion is not done directly in pixel space but in a lower-dimensional latent space, which reduces computational complexity.
Despite \glspl{LDM} provide remarkable results, demonstrated resolutions \cite{avrahami2023blended, rombach2022high} go only up to about $1024\!\times\!1024$ pixels.
Even though the latent space is more compact than the pixel space, increasing the resolution still requires a corresponding enlargement of the latent space.
Therefore, \glspl{LDM} cannot be scaled up arbitrarily.

Another line of methods \cite{saharia2022image, ho2022cascaded, saharia2022photorealistic} generates high-resolution images by passing an initial low-resolution image through a cascade of upscaling diffusion models.
These methods train multiple diffusion models, one for each upscaling stage.
Each of these models takes the full input image of the previous stage as input and predicts an upscaled output.
However, this requires training multiple upscale models, one for each stage.
Also, the last upscaling model must still process the full-resolution image, which is unfeasible for our gigapixel case.

\section{Image generation with diffusion}
\label{sec:diffusion}
Before describing our method in detail, we give the necessary preliminaries about image generation with diffusion.
Diffusion models \cite{pmlr-v37-sohl-dickstein15} generate novel images by pushing noise through a series of denoising steps.
In particular, first, a noise image $\mathbf{x}_0$ is sampled from the Gaussian distribution $\mathcal{N}\left(\mathbf{0}, \sigma_{max}^2 \mathbf{I}\right)$ with variance $\sigma^2_{max}$.
Then, $\mathbf{x}_0$ is sequentially denoised for $N$ steps, producing the sequence $\{\mathbf{x}_i\}_{i\in [0,N]}$, where
the noise level $\sigma_i$ of each $\mathbf{x}_i$ decreases with each step
\begin{equation}
\sigma_0=\sigma_{\text{max}}>\sigma_1>\cdots > \sigma_{min} > \sigma_N=0,
\end{equation}
where $\sigma_{\text{min}}$ is the minimum noise level.
The last image $\mathbf{x}_N$ of this sequential denoising process is noise-free, and follows the data distribution $p_{\text{data}}$ that was used to train the model.

\FloatBarrier
\begin{figure*}[t]
    \centering
    \includegraphics[width=\textwidth]{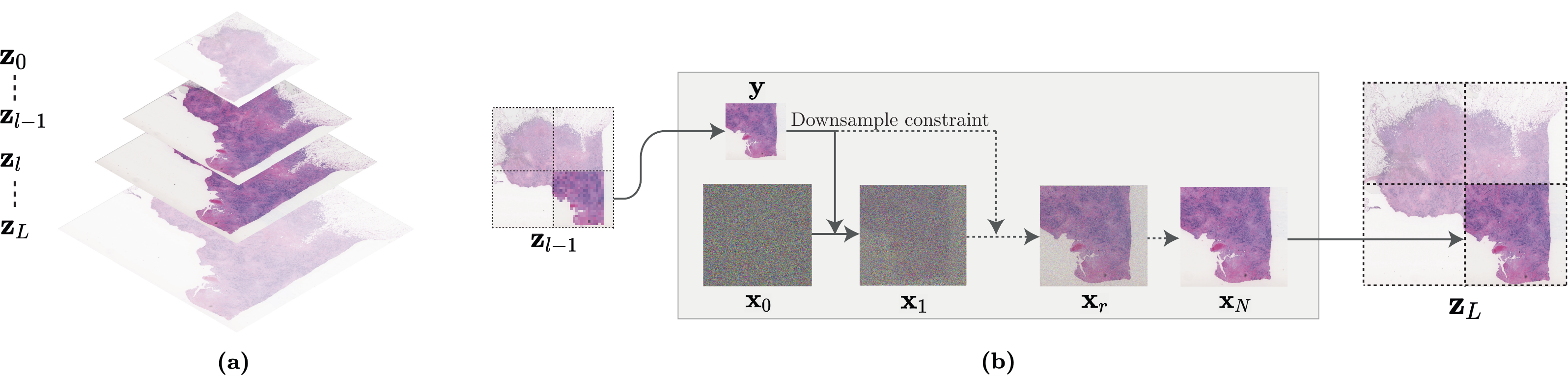}
    \caption{Overview of our method. \textbf{(a)} Shows how we upscale an initial low-resolution image $\mathbf{z}_0$ to a \gls{WSI} $\mathbf{z}_L$ through $L$ upscaling stages. 
    \textbf{(b)} Shows how one stage upscales the image $\mathbf{z}_{l-1}$ to the image $\mathbf{z}_{l}$ using our diffusion-based approach.
    We split the image $\mathbf{z}_{l-1}$ into patches, each having a lower resolution than our diffusion model.
    We then provide each patch as a low-resolution guide $\mathbf{y}$ to a diffusion process.
    Throughout denoising, diffusion is pushed in a direction that satisfies a downsampling constraint with the guide $\mathbf{y}$.
    However, we stop enforcing this constraint after $r$ iterations, which relaxes the constraint.
    Hence, the resulting images $\mathbf{x}_N$ follow the coarse structure of the guide $\mathbf{y}$, with increased resolution and added details.
    Finally, we stitch patches to the image $\mathbf{z}_L$.
    }
    \label{fig:overview}
\end{figure*}

The denoising process of diffusion models can be modelled with \glspl{SDE}.
Additionally, Song \etal \cite{song2020score} proposed that every denoising \gls{SDE} has a correspoding probability flow \gls{ODE} with the same marginals.
While \glspl{SDE} typically converge to higher quality results after numerous steps, \glspl{ODE} can still gives competitive results with significantly fewer steps \cite{song2020score, karras2022elucidating}.
Since our method runs multiple diffusion processes to generate a single \gls{WSI}, we use ODE-based denoising
to keep the overall sampling time within a reasonable scope.

While various variations of the probability flow \gls{ODE} exist, many of them can be expressed with one general equation \cite{karras2022elucidating}:
\begin{equation}
    \mathrm{d} \mathbf{x}=\left[\frac{\dot{s}(t)}{s(t)} \boldsymbol{x}-s(t)^2 \dot{\sigma}(t) \sigma(t) \nabla_{\boldsymbol{x}} \log p\left(\frac{\boldsymbol{x}}{s(t)} ; \sigma(t)\right)\right] \mathrm{d} t,
    \label{eq:pfold}
\end{equation}
where for time $t$ the function $\sigma(t)$ controls the amount of noise, $s(t)$ scales the image, and $\dot{\sigma}(t)$ and $\dot{s}(t)$ are the respective time derivatives.
Setting $\sigma(t)$ and $s(t)$ accordingly, recovers various \glspl{ODE}, \eg, \gls{VP} \cite{song2020score},
variance exploding (VE) \cite{song2020score}, DDIM \cite{song2020denoising}, iDDPM \cite{nichol2021improved} or EDM \cite{karras2022elucidating}.
We use the EDM formulation, 
since it has shown to be favourable in terms of sampling speed and image quality \cite{karras2022elucidating}.
The EDM \gls{ODE} is obtained by setting $s(t)=1$ and the noise-level as $\sigma(t)=t$ in \cref{eq:pfold}. 
For clarity, we continue to denote the noise level as $\sigma(t)$, a function parametrized by time $t$, instead of replacing it directly with $t$, leading to the following EDM \gls{ODE}
\begin{equation}
    \mathrm{d} \mathbf{x}=\left[ - \sigma(t) \,\nabla_{\boldsymbol{x}} \log p\left(\mathbf{x} ; \sigma(t) \right)\right] \mathrm{d} t.
    \label{eq:pfo}
\end{equation}
Following the empirical results and theoretical justifications of Karras \etal \cite{karras2022elucidating}, we set time steps ${t}_{i \in [0,N]}$ as
\begin{equation}
    t_i = \left(\sigma_{\max }{}^{\frac{1}{\rho}}+ \frac{i}{N-1}\left(\sigma_{\min }{}^\frac{1}{\rho}-\sigma_{\max }{}^{\frac{1}{\rho}}\right)\right)^\rho,
\end{equation}
where $\rho$ adjusts between shortening steps near $\sigma_{\text{min}}$ and lengthening those near $\sigma_{\text{max}}$.

To solve the ODE given in \cref{eq:pfo}, one expresses the gradient of the log-likelihood \wrt input $\mathbf{x}$, \ie the score function, as
\begin{equation}
    \nabla_{\mathbf{x}} \log p_{\theta}(\mathbf{x} ; \sigma)= \frac{D_{\theta}(\mathbf{x} ; \sigma)-\mathbf{x}}{\sigma^2},
    \label{eq:score_function}
\end{equation}
where the function $D_{\theta}(\mathbf{x}; \sigma)$ parametrized by $\theta$, takes a noisy image $\mathbf{x}$ and its noise level $\sigma$ as input, and outputs a denoised image.
After training the denoiser $D_{\theta}(\mathbf{x}; \sigma)$,
any numerical ODE solver can be used to solve
the ODE given by putting \cref{eq:score_function} into \cref{eq:pfo}.
Consequently, images can be generated by sampling noise, followed by sequential denoising using the ODE.

\section{Method}
\label{sec:meth}
Our method uses a diffusion model trained on histopathological images of size $M\!\times\!M$ to generate high-resolution WSIs of size $H\!\times\!H$, where $H\!\gg\!M$. 
To generate images of much larger resolution than the resolution of the diffusion model, we use a coarse-to-fine scheme. 
In this scheme, we first sample with the diffusion model an initial image $\mathbf{z}_0\in\mathbb{R}^{M\times M}$. 
Then, we sequentially upscale it in $L$ stages, producing the sequence $\{\mathbf{z}_l\}_{l \in [0,L]}$, where each image $\mathbf{z}_l$ has a $k$-times larger resolution compared to its predecessor $\mathbf{z}_{l-1}$, and the last image $\mathbf{z}_L \in \mathbb{R}^{M \times M}$ resembles a high-resolution \gls{WSI}.
\cref{fig:overview}\,(a) illustrates this coarse-to-fine upscaling.

At each stage $l$ of our coarse-to-fine scheme, we compute the higher-resolution image $\mathbf{z}_l$ through a diffusion process that is guided by 
the preceding lower-resolution image $\mathbf{z}_{l-1}$.
Through this guidance, the image $\mathbf{z}_{l}$ is generated such that it follows the coarse structure of $\mathbf{z}_{l-1}$ while introducing novel details and having increased resolution.
Due to the limited resolution of the diffusion model, we generate $\mathbf{z}_l$ patch-wise.
Importantly, to prevent stitching artefacts in the image $\mathbf{z}_l$, despite patch-wise processing, we introduce a novel technique: grid-shift.
\cref{fig:overview}\,(b) summarizes the upscaling from $\mathbf{z}_{l-1}$ to $\mathbf{z}_l$.

In the following, we describe our method in detail.
We start with the design of our diffusion denoising function in \cref{sec:denoiser} and its training in \cref{sec:training}.
Followed by our guided denoising step for diffusion in \cref{sec:sampling} and the description of grid-shift in \cref{sec:grid}.

\subsection{Diffusion denoiser}
\label{sec:denoiser}
As discussed in \cref{sec:diffusion}, for diffusion, we need a denoiser function $D_{\theta}(\mathbf{x}; \sigma)$
that denoises images at each timestep.
We propose to condition the denoiser $D_{\theta}(\mathbf{x}; \sigma)$ not only with noise level $\mathbf{\sigma}$ but also with the spatial image resolution s in $\SI{}{\micro \metre} / \SI{}{\px}$.
While in many applications the spatial resolution is unknown, it is consistently available in our case, as slide scanners usually save it in the metadata of \glspl{WSI}.
Conditioning allows us to control the spatial resolution of generated images.
This is crucial for our coarse-to-fine scheme. 
Setting a high spatial resolution for the initial image ensures it depicts a macroscopic overview of a tissue sample. 
While decreasing spatial resolution accordingly at later refinement stages,
conditions the network to introduce small details like cellular structures.

For denoising, we introduce a network $F_{\theta}(\mathbf{x}; \sigma, s)$,
where we implement the conditioning on noise $\sigma$ and spatial resolution $s$ with a sinusoidal positional encoding \cite{vaswani2017attention}.
However, we do not use the network $F_{\theta}(\mathbf{x}; \sigma, s)$ to directly denoise images, \ie ${D_{\theta}(\mathbf{x}; \sigma, s) = F_{\theta}(\cdot)}$.
Instead, we use the network preconditioning of Karras \etal \cite{karras2022elucidating}
\begin{equation}
    D_\theta(\mathbf{x} ; \sigma, s)=c_{\text {skip }}(\sigma) \, \mathbf{x}+c_{\text {out }}(\sigma) \, F_\theta\big(c_{\text {in }}(\sigma) \, \mathbf{x} ; \sigma, s \big),
    \label{eq:denoiser_skipped}
\end{equation}
where the functions $c_{\text {in}}(\sigma)$ and $c_{\text {out }}(\sigma)$ scale the inputs and outputs of the network $F_{\theta}(\mathbf{x}; \sigma, s)$,
and $c_{\text {skip }}(\sigma)$ is a $\sigma$-dependent skip connection.
These three functions scale network input and training targets to unit variance across all noise levels $\sigma$,
which is beneficial for neural network training \cite{huang2023normalization}.
Additionally, $c_{\text {skip }}(\sigma)$ controls, if for denoising, the network has to
predict the denoised image directly, only the noise component or a mixture of both.
Empirically, it has been demonstrated that it depends on the noise level $\sigma$ which of these cases is easier to learn,
and $c_{\text {skip }}(\sigma)$ is set to adapt accordingly.
We provide the full expressions of $c_{\text {in}}(\sigma)$, $c_{\text {out}}(\sigma)$ and $c_{\text {skip }}(\sigma)$ in the appendix.

\subsection{Training}
\label{sec:training}
Using our denoiser function $D_\theta(\mathbf{x} ; \sigma, s)$ given in \cref{eq:denoiser_skipped}, we can define the training loss for the diffusion model.
In particular, we minimize the expected $L_2$ denoising error
\begin{equation}
    \mathbb{E}_{s, \tilde{\mathbf{x}}, \sigma, \mathbf{n}}\big[\lambda(\sigma) \|D_{\theta}(\tilde{\mathbf{x}} + \mathbf{n} ; \sigma, s) - \tilde{\mathbf{x}} \|_2^2 \big],
    \label{eq:loss}
\end{equation}
where the function $\lambda(\sigma)$ weights loss terms equally across all noise levels $\sigma$.
At first, we sample the spatial resolution uniformly ${s \sim \unif(s_{\text{min}}, s_{\text{max}})}$, where $s_{\text{min}}$ and $s_{\text{max}}$ refer to the smallest
respectively largest spatial resolution of image patches in the training dataset.
Then, we sample images from the distribution of training patches having spatial resolution $s$, \ie $\tilde{\mathbf{x}} \sim p_{\text{train}\vert s}$.
Finally, we sample noise levels $\sigma$ from a log-normal distribution, and noise as $\mathbf{n} \sim \mathcal{N}\left(\mathbf{0}, \sigma^2 \mathbf{I}\right)$.


\subsection{Guided denoising step}
\label{sec:sampling}
Like a conventional diffusion denoising step, our guided denoising step removes noise from a noisy input image $\mathbf{x}_i$ with noise level $\sigma(t_i)$ such that the result $\mathbf{x}_{i+1}$ has noise level $\sigma(t_{i+1}) < \sigma(t_i)$. 
Additionally, we guide the denoising step with a low-resolution guidance patch ${\mathbf{y} \in \mathbb{R}^{d \times 1}}$ from the preceding layer $\mathbf{z}_{l-1}$.
The goal of guidance is that the fully denoised image $\mathbf{x}_0$ follows the coarse structure of the guidance patch $\mathbf{y}$ while having additional details and a higher resolution.
We implement this guidance through a relaxed super-resolution constraint.

For further derivations,
we denote ${\mathbf{u} \in \mathbb{R}^{D\times 1}}$ as the output of the denoiser function $D_\theta(\mathbf{x}_i ; \sigma(t_i), s)$ at step $t_i$.
Notably, ${\mathbf{u}}$ gives at each denoising step an estimate of the \textit{fully} denoised image $\mathbf{x}_0$.
In our guided denoising step, we replace the initial estimation $\mathbf{u}$ of the denoised image with a guided estimate $\bar{\mathbf{u}}$, which is computed to be close to $\mathbf{u}$ while additionally satisfying a guidance constraint.
This basically resembles the concept of projected gradient descent.

For guidance, we introduce the downsampling constraint $\mathbf{A}{\mathbf{u}}=\mathbf{y}$, where $\mathbf{A} \in \mathbb{R}^{d \times D}$ is a known linear downsampling operator.
Therefore, downsampling the estimate $\mathbf{u}$ should equal the low-resolution guide $\mathbf{y}$. 
We can compute the guided estimate $\bar{\mathbf{u}}$ through the following optimization problem

\begin{equation}
    \bar{\mathbf{u}} = \argmin_{\bar{\mathbf{u}}} \frac{1}{2}\|\mathbf{u}-\bar{\mathbf{u}}\|^2 \quad \text { s.t. } \mathbf{A}\bar{\mathbf{u}} = \mathbf{y},
    \label{eq:optimization}
  \end{equation}
  that can solved using the method of Lagrangian multipliers.
  We provide a full derivation in the appendix and continue here with the solution
  \begin{equation}
     \bar{\mathbf{u}} = (\mathbf{I} - \mathbf{A}^\dagger \mathbf{A}) \mathbf{u} + \mathbf{A}^{\dagger} \mathbf{y},
     \label{eq:soll}
  \end{equation}
 where $\mathbf{A}^{\dagger}$ is the pseudoinverse for full row rank matrices
  \begin{equation}
    \mathbf{A}^{\dagger} = \mathbf{A}^T(\mathbf{A} \mathbf{A}^T)^{-1}.
  \end{equation}
  Notably, \cref{eq:soll} resembles the proposed rectification equation of DDNM \cite{wang2022zero}, a method to solve linear inverse problems with diffusion models.
  However, DDNM presents a different derivation based on a range-space null-space decomposition.
  Also, DDNM uses SDE-based diffusion processes, contrary to our ODE-based setting, leading to a different application of \cref{eq:soll}.

So far, our guidance resembles unrelaxed super-resolution.
However, we do not strictly enforce the downsampling constraint, but relax it. 
Hence, we allow slight differences, between the downsampled fully denoised image $\mathbf{x}_0$ and the low-resolution guide $\mathbf{y}$.
For relaxation, we stop replacing the estimate $\mathbf{u}$ with the guided estimate $\bar{\mathbf{u}}$ at iterations $i$ where $i>r$.
Consequently, in the last denoising steps, changes to the image are allowed that do not satisfy the downsample constraint.
The strength of relaxation is controlled through $r$. If $r=0$, the guidance constraint is enforced at all iterations, leading to no relaxation. 
Contrarily, if $r=N$, the constraint is never applied, leading to full relaxation. 
By setting $r$ to values in between controls the amount of relaxation accordingly.

There are multiple reasons why we relax the downsample constraint.
In our coarse-to-fine scheme, we do not pursue strict upsampling; instead, the diffusion model should add new details at every stage.
Adhering strictly to the downsample constraint would restrict the flexibility to add new details.
Furthermore, without relaxation, the downsampling constraint would be enforced across all upscaling stages.
This is unreasonable due to the vast upscaling factors we face.
For instance, if we have a diffusion model with input size $512\!\times\!512$ and generate a \gls{WSI} with a resolution of $\numprint{65536}\!\times\!\numprint{65536}$, we have an upscaling factor of $128$. 
Consequently, for a $512\!\times\!512$ area in the full-resolution WSI, the downsampling constraint would be enforced with a $4\!\times\!4$ patch in the lowest-resolution image. 
Clearly, this does not introduce any meaningful information.
Moreover, without relaxation, even single-pixel errors at the lowest-resolution can distort large areas in the full-resolution image.

\begin{algorithm}[t]
    \small
    \caption{Guided denoising step}
    \begin{algorithmic}[1]
        \Require{Noisy image $\mathbf{x}_i$, guide $\mathbf{y}$, step $i$, spatial-resolution $s$}
        \Ensure{Denoised image $\mathbf{x}_{i+1}$}
        \State $\mathbf{u} \leftarrow D_\theta(\mathbf{x}_i ; \sigma(t_i), s)$
        \If{$i < r$}
            \State $\bar{\mathbf{u}} \leftarrow \left(\mathbf{I}-\mathbf{A}^{\dagger} \mathbf{A}\right) \mathbf{u} + \mathbf{A}^{\dagger} \mathbf{y} $
            \State $\mathbf{d}_i \leftarrow \left(\mathbf{x}_i - \bar{\mathbf{u}}\right) / \sigma(t_i)$
        \Else
            \State $\mathbf{d}_i \leftarrow \left(\mathbf{x}_i - \mathbf{u}\right) / \sigma(t_i)$
        \EndIf
        \State $\mathbf{x}_{i+1} \leftarrow \mathbf{x}_i + \left(t_{i+1} - t_i\right) \mathbf{d}_i$
        \If{$t_{i+1} \neq 0$}  \Comment{Skip $2^{nd}$ order correction at last step}
        \State $\mathbf{u}^{\prime} \leftarrow D_\theta(\mathbf{x}_{i+1} ; \sigma(t_{i+1}), s)$
        \If{$i < r$}
            \State $\bar{\mathbf{u}}^{\prime} \leftarrow \left(\mathbf{I}-\mathbf{A}^{\dagger} \mathbf{A}\right) \mathbf{u}^{\prime} + \mathbf{A}^{\dagger} \mathbf{y} $
            \State $\mathbf{d}_i^{\prime} \leftarrow \left(\mathbf{x}_{i+1} - \bar{\mathbf{u}}^{\prime} \right) / \sigma(t_{i+1})$
        \Else
            \State $\mathbf{d}_i^{\prime} \leftarrow \left(\mathbf{x}_{i+1} - {\mathbf{u}}^{\prime} \right) / \sigma(t_{i+1})$
        \EndIf
            \State $\mathbf{x}_{i+1} \leftarrow \mathbf{x}_i + \left(t_{i+1}  - t_i\right)\left(\frac{1}{2} \mathbf{d}_i + \frac{1}{2} \mathbf{d}_i^{\prime}\right)$
        \EndIf
        \State \Return $\mathbf{x}_{i+1}$
    \end{algorithmic}
    \label{alg:heun}
\end{algorithm}

Finally, with \cref{eq:score_function} the score function, \cref{eq:pfo} the EDM \gls{ODE}, and our guided estimation $\bar{\mathbf{u}}$, inplace of the 
denoiser function $D_\theta(\mathbf{x} ; \sigma, s)$, we get
\begin{equation}
\mathrm{d} \mathbf{x} = \frac{\mathbf{x} - \bar{\mathbf{u}}}{\sigma(t)}\, \mathrm{d} t.
\label{eq:ours}
\end{equation}
In principle, we can solve \cref{eq:ours} with any black-box \gls{ODE} solver.
Here, we use Heun's 2nd order solver \cite{ascher1998computer}, a predictor-corrector method, 
which has shown a good tradeoff between truncation error and number of function evaluations in the context of diffusion models \cite{jolicoeur2021gotta}.
\cref{alg:heun} summarizes our guided denoising step.
Note that skipping lines $8$ to $15$ simplifies Heun's 2nd order method to a simple Euler step.

\subsection{Grid-shift}
\label{sec:grid}
To avoid stitching artefacts in our patch-wise refinement scheme, we propose grid-shift. 
If we simply do patch-wise refinement and then stitch the refined patches back to a high-resolution image, the result could suffer from stitching artefacts. 
This is because there is no guarantee that the areas at the edges of neighbouring patches align such that they can be stitched seamlessly.

A recently proposed method to avoid stitching artefacts at patch-wise image processing with diffusion models is mask-shifting \cite{wang2022zero}. 
The idea of mask-shifting is to use overlapping patches. For each patch, areas that overlap with previously computed neighbouring patches are held constant during diffusion. 
This incorporates the content of a previously computed patch into the computation of its following patches. 
And consequently leads to smooth transitions between neighbouring patches.

\begin{algorithm}[t]
    \small
    \caption {Coarse-to-fine scheme with grid-shift}
    \begin{algorithmic}[1]
        \Require{Low-resolution image $\mathbf{z}_0$, and its spatial-resolution $s$}
        \Ensure{High-resolution \gls{WSI} $\mathbf{z}_{L}$}
        \For{Stage $l$ in $[1, L]$}
            \State  $s \leftarrow s / k$ \Comment{Adapt spatial-resolution to current stage}
            \State $\boldsymbol{x}_{0} \sim \mathcal{N}\left(\mathbf{0} | \sigma_{\text{max}}^2 \mathbf{I}\right)$
            \For{$i$ in $[1, \ldots, N]$}
                \State shift\_patch\_grid()
                \For{$\mathbf{x}$, $\mathbf{y}$ in $\text{patch}(\mathbf{x}_{i-1})$, $\text{patch}(\mathbf{z}_{l-1})$}
                    \State $\mathbf{x}_{i,p} \leftarrow \text{\cref{alg:heun}}(\mathbf{x}, \mathbf{y}, i, s)$
                \EndFor
            \State $\mathbf{x}_{i} \leftarrow \text{stitch\_patches}([\mathbf{x}_{i,0}, \dots, \mathbf{x}_{i,P}])$
            \EndFor
            \State $\mathbf{z}_{l} \leftarrow \mathbf{x}_{N}$
        \EndFor
        \State \Return $\mathbf{z}_L$
    \end{algorithmic}
    \label{alg:grid_shift_sampling}
\end{algorithm}

\begin{figure}[t]
    \centering
    \includegraphics[width=.45\textwidth]{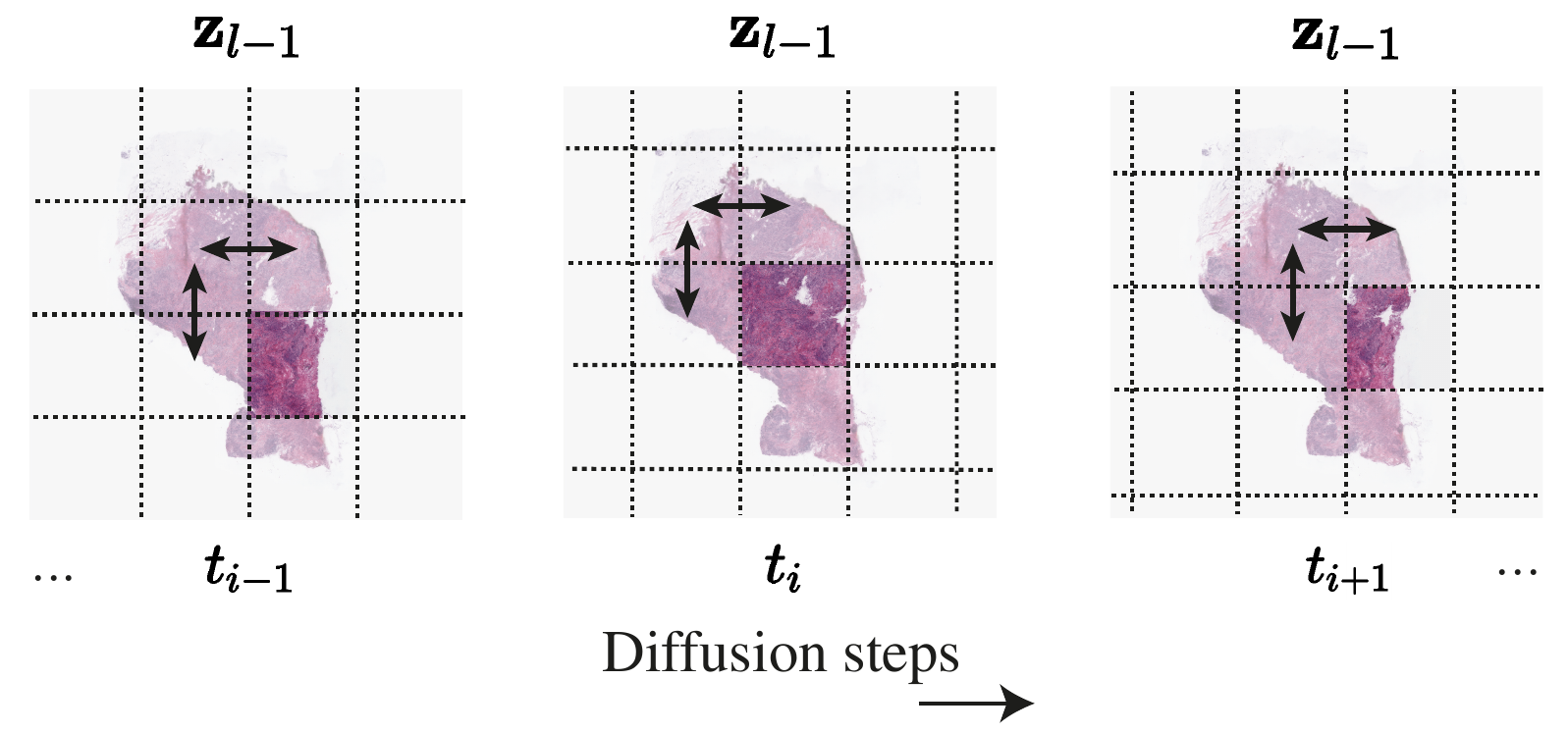}
    \vspace{-.9em}
    \caption{Visualization of grid-shift. 
    After each diffusion step, we shift the patch grid that is used to extract guidance patches from the preceding image $\mathbf{z}_{l-1}$.
    }
    \label{fig:grid-shift}
\end{figure}

We argue that mask-shifting has two drawbacks.
At first, patches must be processed sequentially, which is not trivial to parallelize.
And second, using overlapping patches increases the amount of total patches to process, increasing the computation time significantly depending on the amount of overlap between patches.

With grid-shift, we address both discussed drawbacks of mask-shifting by interleaving diffusion iterations with patching. 
Instead of using a fixed grid to extract patches, we shift the patch grid after each diffusion step. 
This makes patch boundaries temporary since they change after each diffusion step. 
Consequently, information between neighbouring patches is continuously transferred, resulting in a more coherent result without visible seams. 
\cref{fig:grid-shift} illustrates grid-shift. 
In our experiments, we shifted the patch-grid with random translations and padded boundary patches with the background colour.

Grid-shift has two computational advantages over mask-shifting. 
First, it does not increase the total amount of patches to process. 
And second, during one diffusion step, all patches are processed independently. 
Therefore, grid-shift is trivial to parallelize, \eg, for a multi-GPU implementation.
\cref{alg:grid_shift_sampling} shows our full coarse-to-fine scheme with grid-shift.

\section{Experiments}
To evaluate our method, we performed a user study with pathologists, and quantitative evaluations. 
Particularly, quantitative evaluation is challenging due to the lack of a suitable standardized metric.
Common metrics for generative models, such as FID \cite{szegedy2016rethinking}, or improved precision (IP) and improved recall (IR) \cite{kynkaanniemi2019improved} require features from pre-trained networks. 
These standardized networks have a fixed input size of $224\!\times\!224$. 
Downscaling \glspl{WSI} to this resolution would discard most information, making the metrics inconclusive. 
Also, these metrics require large sample sizes of $\numprint{50000}$ images to provide consistent results. 
Generating that many \glspl{WSI} is infeasible, considering that we need $\sim\!40$ minutes for a single \gls{WSI}. 
Moreover, the metrics utilize feature spaces strongly influenced by ImageNet classes \cite{Kynkaanniemi2022}.
Using these feature spaces to evaluate images from entirely different domains than ImageNet, such as histopathology images might be problematic.

Due to the discussed limitations, our quantitative evaluations are restricted to isolated evaluations of our diffusion model without the coarse-to-fine scheme.
In terms of metrics, we use IP and IR, following Moghadam \etal \cite{moghadam2023morphology}.
However, we add that these metrics should be taken with reservations due to their ImageNet-related feature spaces.

Additionally, to the experiments presented in this section, we compare our method with multiple super-resolution methods in the appendix.

\paragraph{Data.}
For all experiments, we used the The Cancer Genome Atlas Breast Invasive Carcinoma (TCGA-BRCA) dataset \cite{weinstein2013cancer}. 
The dataset contains \numprint{1978} high-resolution WSIs stained using various protocols showing diverse tissue types, \eg, epithelium, muscle, and connective tissue.
For training, we extract patches from the dataset with spatial resolutions ranging from $s_{\text{min}}=0.3\, \SI{}{\micro \metre} / \SI{}{\px}$ to ${s_{\text{max}}=150\, \SI{}{\micro \metre} / \SI{}{\px}}$.


\paragraph{Setup.}
We generate WSIs at a resolution of ${\numprint{65536}\!\times\!\numprint{65536}}$ pixels.
For the diffusion model, we use a resolution of $512\!\times\!512$ pixels. 
In our coarse-to-fine scheme, we use an upscaling factor of $k=2$ at each stage, resulting in $L=7$ stages in total. 
Initial images $\mathbf{z}_0$ are generated with a spatial-resolution randomly between 80 $\SI{}{\micro \metre} / \SI{}{\px}$ and 150 $\SI{}{\micro \metre} / \SI{}{\px}$.
We set the number of diffusion denoising steps to $N=40$ based on the results of \cref{sec:sovler}.
The relaxation parameter of our relaxed super-resolution constraint is set to $r=28$, which was manually tuned towards a good tradeoff between consistency and novelty.
For the downsampling operator $\mathbf{A}$ we use average-pooling.

We train for five days on four NVIDIA Quadro RTX 8000 GPUs with 48 GB of memory each.
It took on average $\sim\!\!40$ minutes on one GPU to sample a single \glspl{WSI} with a resolution of $\numprint{65536}\!\times\!\numprint{65536}$ pixels.
For the diffusion-related hyperparameters, we use, if not otherwise stated, the proposed settings of Karras \etal \cite{karras2022elucidating}.
Likely, an extensive hyperparameter search could further improve our results, but given the extensive training cost, it is beyond our computational capacities.


\begin{figure}[t]
    \centering
    \includegraphics[width=.48\textwidth]{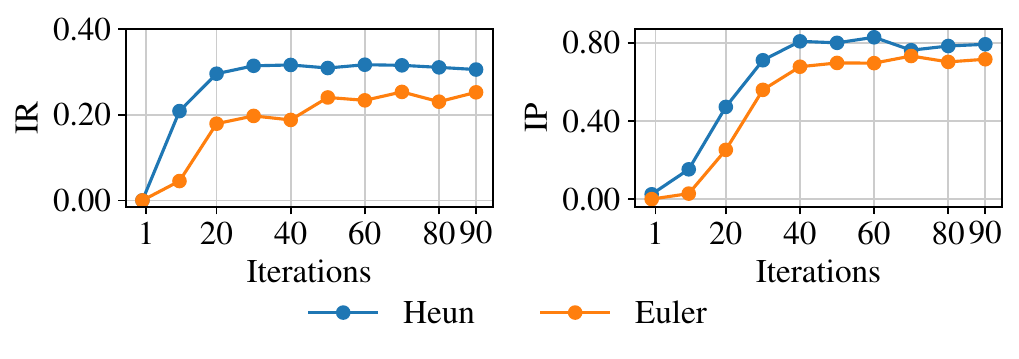}
    \vspace{-2.5em}
    \caption{IR and IP values for $512\!\times\!512$ patches with $1\SI{}{\micro \metre} / \SI{}{\px}$ for different numbers of denoising iterations using Heun and Euler as \gls{ODE} solver.}
    \label{fig:sampler_comp}
\end{figure}

\subsection{Number of diffusion iterations}
An important hyperparameter we must choose is the number of diffusion iterations $N$. 
Too few iterations degrade image quality, while too many might increase runtime unnecessarily. 
Finding the right balance is crucial for us since we have to run many diffusion processes to sample a single \gls{WSI}.
To this end, we compute IR and IP scores for generating $512\!\times\!512$ patches with a spatial resolution of $1\SI{}{\micro \metre} / \SI{}{\px}$ across different iteration numbers $N$.
We also validated if using Heun's 2nd order method is beneficial over a plain Euler solver. 
\cref{fig:sampler_comp} shows the results. 
According to the metrics, the Heun solver showed preferable performance. 
After an additional manual inspection, we chose $N=40$ as a good tradeoff between image quality and runtime for further experiments.
\label{sec:sovler}

\subsection{Image quality across spatial resolutions}
To evalute our spatial resolution conditioning of the model, we compute IP and IR metrics across a variety of different spatial resolutions.
We obtained all results from a $\textit{single}$ model trained with uniformly sampled spatial resolutions as described in our training setup.
We then conditionally sampled $\numprint{50000}$ images for each spatial resolution, and compared them with images of identical spatial resolution from the training dataset.
\cref{table:my_label} shows the result.
According to the metrics, performance is relatively consistent across the full range of spatial resolutions without any major outliers.
    \begin{table}[t]
    \footnotesize
        \centering
        \begin{tabular}{llllllllc}
        \toprule
        & \multicolumn{8}{c}{Spatial Resolution [\SI{}{\micro \metre} / \SI{}{\px}]} \\
        \cmidrule(lr){2-8}
        & 0.3 & 1.0 & 25 & 50 & 100 & 150 & ${\unif(0.3, 150)}$ \\
        \midrule
        IP & 0.81 & 0.82 & 0.82 & 0.85 & 0.82 & 0.84 & 0.86 & \\
        IR & 0.32 & 0.33 & 0.36 & 0.38 & 0.37 & 0.36 & 0.34 \\
        \bottomrule
        \end{tabular}
        \caption{IP and IR for $512\!\times\!512$ patches at varying spatial resolutions.
        The last column shows results for uniformly sampled spatial resolutions between $0.3$ and $150$.}
        \label{table:my_label}
    \end{table}



\begin{figure}[t]
    \centering
    \begin{tabular}{c@{}@{}c@{\hspace{0.05cm}}c@{\hspace{-0.04cm}}c@{}}

        \rotatebox{90}{Nearest neighbor} & 
        \begin{subfigure}[b]{0.147\textwidth}
            \includegraphics[width=\textwidth]{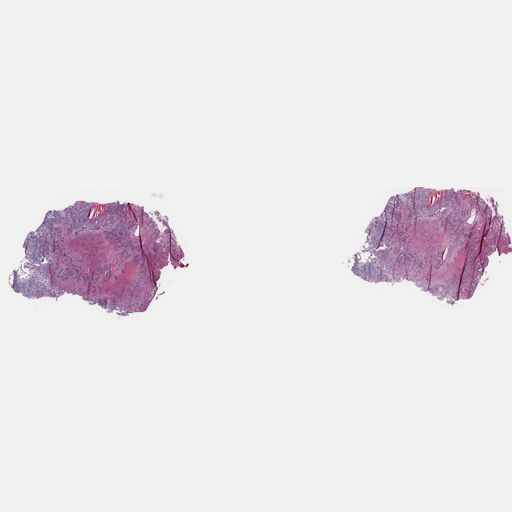}
        \end{subfigure} &
        \begin{subfigure}[b]{0.147\textwidth}
            \includegraphics[width=\textwidth]{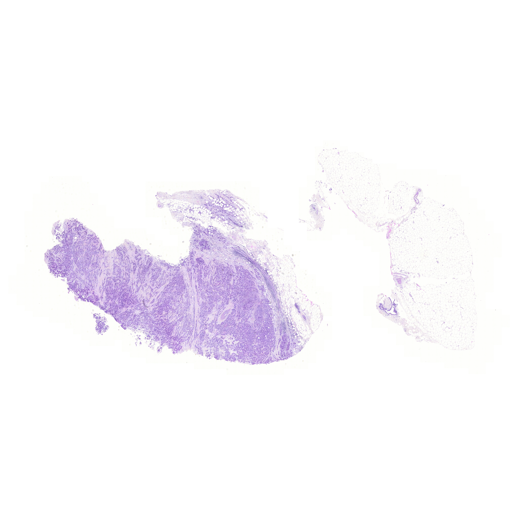}
        \end{subfigure} &
            \hspace{0.00cm}
        \begin{subfigure}[b]{0.147\textwidth}
            \includegraphics[width=\textwidth]{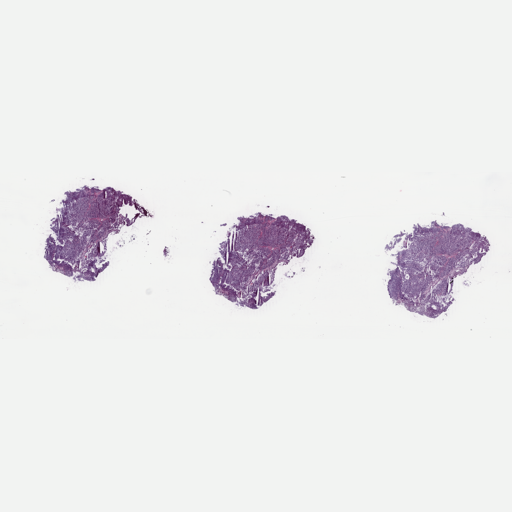}
        \end{subfigure}
        \vspace{0.15cm}
        \\
        \vspace{-0.05cm}

        {\rotatebox{90}{$\hspace{1.0cm}1\times$}} &
        \begin{subfigure}[b]{0.147\textwidth}
            \includegraphics[width=\textwidth]{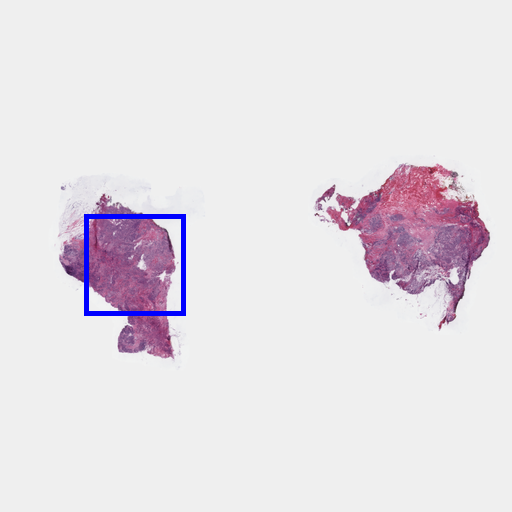}
        \end{subfigure} &
        \begin{subfigure}[b]{0.147\textwidth}
            \includegraphics[width=\textwidth]{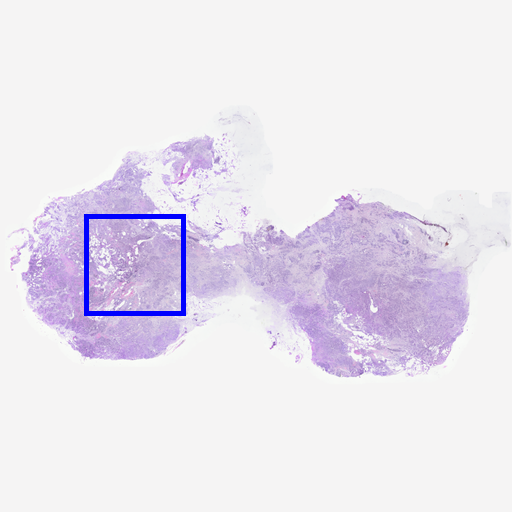}
        \end{subfigure} &
        \begin{subfigure}[b]{0.147\textwidth}
            \includegraphics[width=\textwidth]{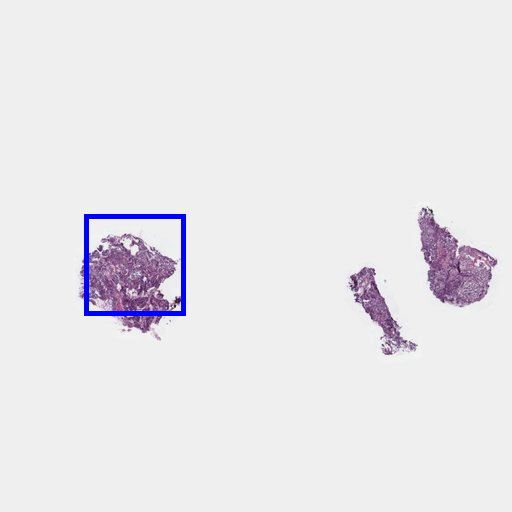}
        \end{subfigure} \\
        \vspace{-0.05cm}

        \rotatebox{90}{$\hspace{1.0cm}5\times$} & 
        \begin{subfigure}[b]{0.147\textwidth}
            \includegraphics[width=\textwidth]{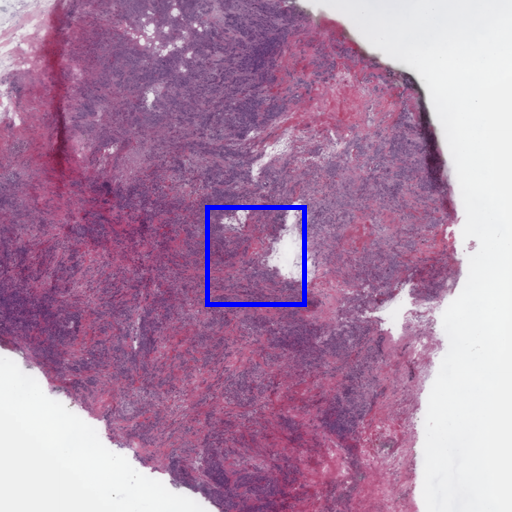}
        \end{subfigure} &
        \begin{subfigure}[b]{0.147\textwidth}
            \includegraphics[width=\textwidth]{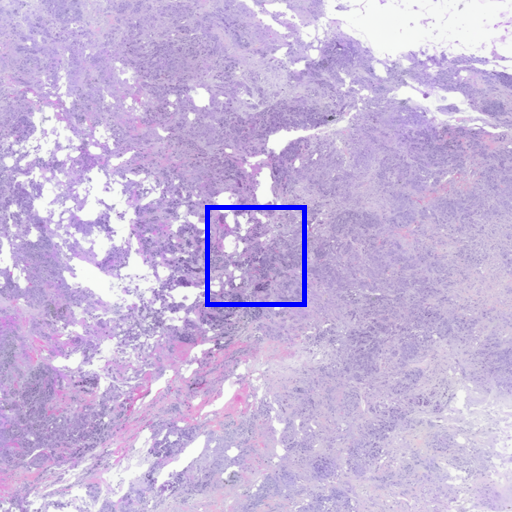}
        \end{subfigure} &
        \begin{subfigure}[b]{0.147\textwidth}
            \includegraphics[width=\textwidth]{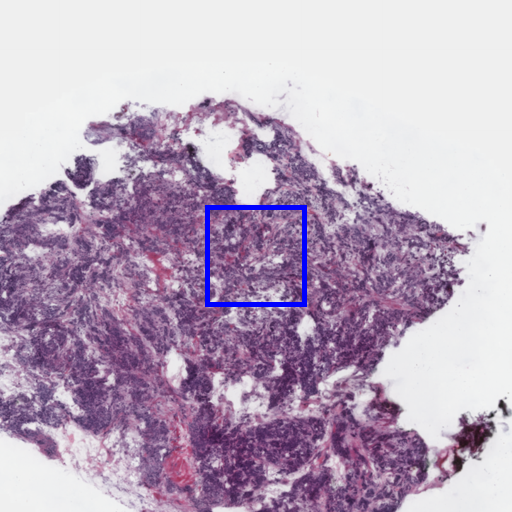}
        \end{subfigure} \\
        \vspace{-0.05cm}

        \rotatebox{90}{$\hspace{1.cm}25 \times$} & 
        \begin{subfigure}[b]{0.147\textwidth}
            \includegraphics[width=\textwidth]{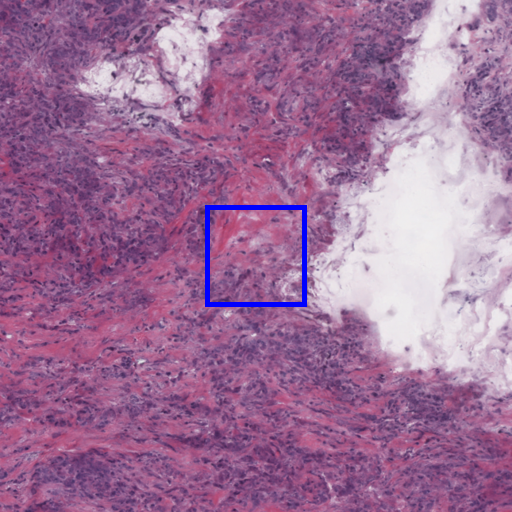}
        \end{subfigure} &
        \begin{subfigure}[b]{0.147\textwidth}
            \includegraphics[width=\textwidth]{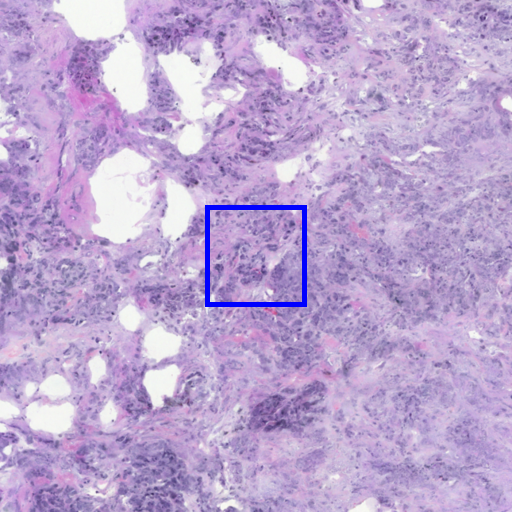}
        \end{subfigure} &
        \begin{subfigure}[b]{0.147\textwidth}
            \includegraphics[width=\textwidth]{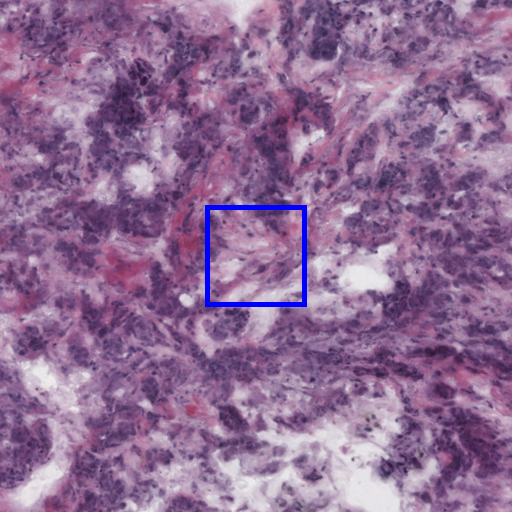}
        \end{subfigure} \\
        \vspace{-0.05cm}

        \rotatebox{90}{$\hspace{.9cm}128 \times$} & 
        \begin{subfigure}[b]{0.147\textwidth}
            \includegraphics[width=\textwidth]{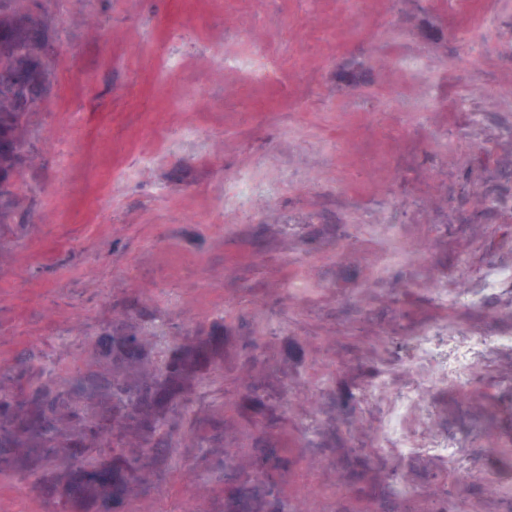}
        \end{subfigure} &
        \begin{subfigure}[b]{0.147\textwidth}
            \includegraphics[width=\textwidth]{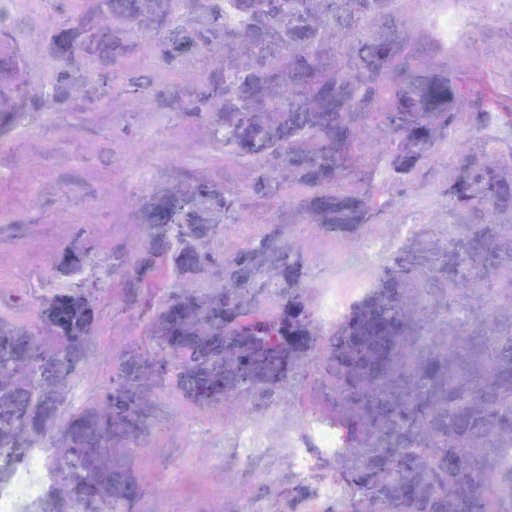}
        \end{subfigure} &
        \begin{subfigure}[b]{0.147\textwidth}
            \includegraphics[width=\textwidth]{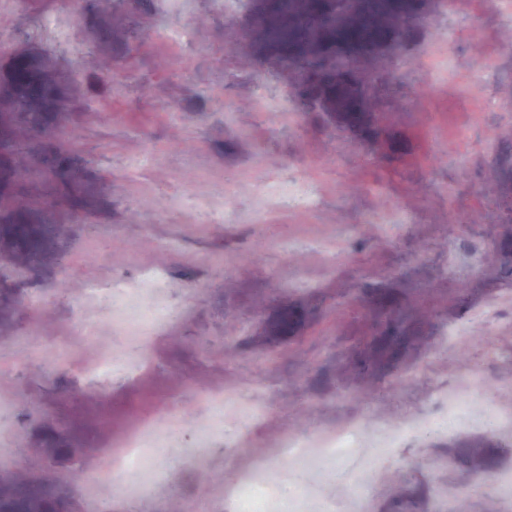}
        \end{subfigure} \\
    \end{tabular}
    \caption{\glspl{WSI} with a resolution of $\numprint{65536}\!\times\!\numprint{65536}$ pixels generated by our method.
    We show $\numprint{512}\!\times\!\numprint{512}$ patches extracted at different magnifications, the blue rectangle shows the location of the patch in the subsequent row.
    The top row shows for each \gls{WSI} the nearest neighbor in the training data.
    To find neighbors, we resized \glspl{WSI} to $\numprint{512}\!\times\!\numprint{512}$ and compared \glspl{WSI} in the feature space of Inception-v3.
    }
    \label{fig:qual}
\end{figure}

\subsection{Relaxation parameter}
\begin{figure}[t]
    \centering
    \begin{subfigure}{0.2\columnwidth}
        \includegraphics[width=\textwidth]{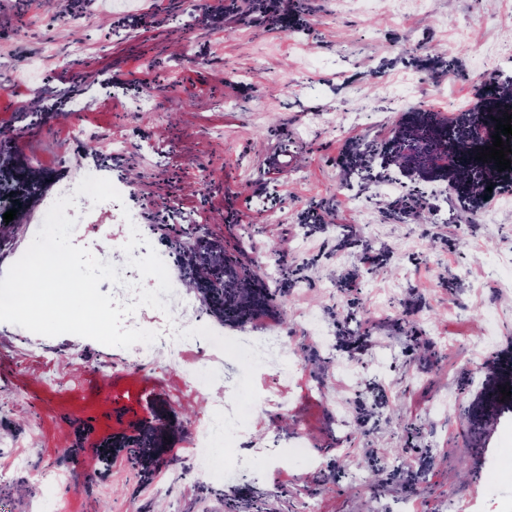}
        \caption*{Initial image}
    \end{subfigure}%
    \hfill
    \begin{subfigure}{0.2\columnwidth}
        \includegraphics[width=\textwidth]{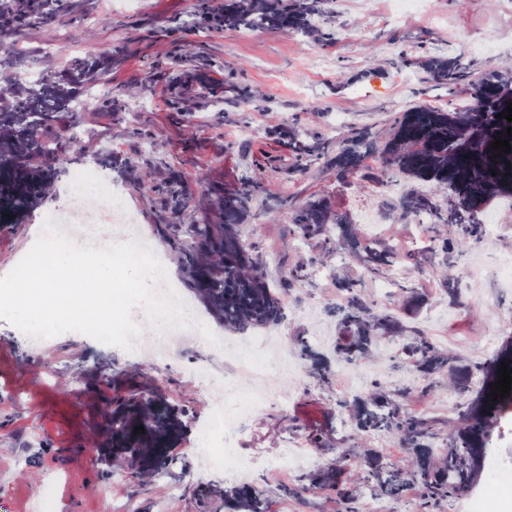}
        \caption*{$r=20$}
    \end{subfigure}%
    \hfill
    \begin{subfigure}{0.2\columnwidth}
        \includegraphics[width=\textwidth]{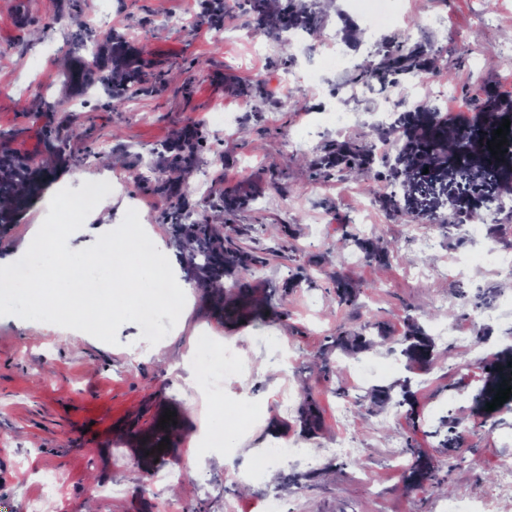}
        \caption*{$r=16$}
    \end{subfigure}%
    \hfill
    \begin{subfigure}{0.2\columnwidth}
        \includegraphics[width=\textwidth]{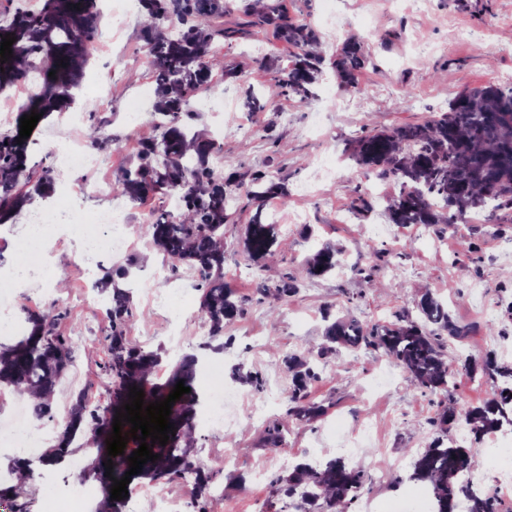}
        \caption*{$r=12$}
    \end{subfigure}%
    \hfill
    \vspace{-.5em}
    \caption{Decreasing the parameter $r$ relaxes the super-resolution constraint.}
    \vspace{-1.0em}
    \label{fig:cec}
\end{figure}
To evaluate the influence of our super-resolution relaxation parameter $r$, we perform a simple experiment.
We sample a $512\!\times\!512$ sized image with our diffusion model, downsample it to $256\!\times\!256$, and provide it as a guide $\mathbf{y}$ to a diffusion process guided by our relaxed super-resolution constraint.
We repeat this for multiple values of the relaxation parameter $r$.
\cref{fig:cec} shows the result, it can be clearly seen how consistency with the initial images decreases with decreasing relaxation parameter $r$.

\subsection{User study}
To evaluate the quality of our synthetic WSIs, we conducted a user study with three pathologists.
For the study, we used $20$ synthetic WSIs, and $20$ real WSIs randomly chosen from the training data. 
We presented the WSIs to the pathologists in random order and asked them to identify whether each WSI was synthetic or real. 
To this end, they were given a slider to select values between $0$ = "I believe the slide is real." and $100$ = "I believe the slide is synthetic.". 
Values in between represented corresponding gradations between the two extremes. 
\cref{fig:qual} shows three \glspl{WSI} that were part of the study.

\begin{figure}[t]
    \centering
    \hspace{-1.em}
    \includegraphics[width=.49\textwidth]{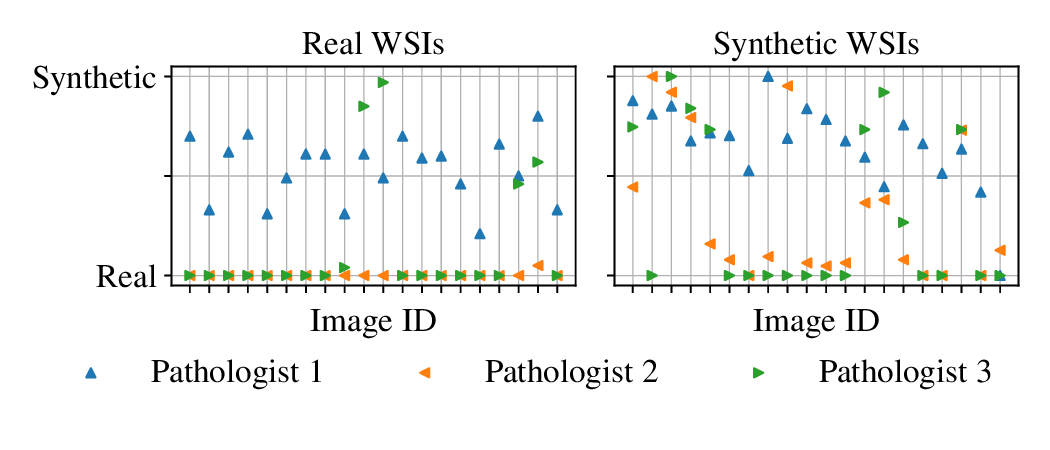}
    \vspace{-2.8em}
    \caption{Result of our user-study. The plots show how three pathologists rated the realness of 20 real versus 20 synthetic \glspl{WSI}.
    }
    \label{fig:study}
\end{figure}

\cref{fig:study} shows the ratings for all individual images.
Despite our study's limited sample size, our primary goal was to assess whether our method could generate plausible-looking WSIs.
The results of our study indicate that this is the case, as pathologists could not consistently distinguish our synthetic WSIs from real ones.

\section{Conclusion}
We presented a method that generates synthetic histopathological WSIs at resolutions up to $\numprint{65536}\!\times\!\numprint{65536}$.
We evaluated parts of our method quantitatively and also performed a user study with pathologists. 
Our study's results showed that pathologists could not consistently differentiate the WSIs generated by our method from real ones.
In the future, the duration of WSI generation could be further reduced by incorporating distillation-based diffusion models \cite{song2023consistency, salimans2022progressive}.

\begin{footnotesize}
\textbf{Acknowledgement}
\quad This work has been co-funded by the Austrian Science Fund (FWF), Project: P-32554 explainable Artificial Intelligence.
\end{footnotesize}

{\small
\bibliographystyle{ieee_fullname}
\bibliography{PaperForReview}
}

\begin{appendix}

\twocolumn[ 
   \centering
   \Large\bfseries {Diffusion-based generation of Histopathological Whole Slide Images at a Gigapixel scale} \\
   \vspace{.8em}
    \mdseries \Large Supplementary material
   \vspace{.8em}

   \small
   \author{Robert Harb\textsuperscript{1,2}, Thomas Pock\textsuperscript{1}, Heimo Müller\textsuperscript{2}\\
   \textsuperscript{1}Institute of Computer Graphics and Vision, Graz University of Technology, Austria\\
   \textsuperscript{2}Diagnostic and Research Institute of Pathology, Medical University of Graz, Austria\\
   {\tt\small \{robert.harb, pock\}@icg.tugraz.at, heimo.mueller@medunigraz.at}
   }
   \vspace{2em} 
]
    \noindent

    \crefalias{section}{appsec}

\begin{figure*}[ht]
    \hspace{1.0em}
    \begin{tabular}{
      >{\centering\arraybackslash}p{\dimexpr.132\textwidth-2\tabcolsep}
      >{\centering\arraybackslash}p{\dimexpr.14\textwidth-2\tabcolsep}
      >{\centering\arraybackslash}p{\dimexpr.14\textwidth-2\tabcolsep}
      >{\centering\arraybackslash}p{\dimexpr.14\textwidth-2\tabcolsep}
      >{\centering\arraybackslash}p{\dimexpr.14\textwidth-2\tabcolsep}
      >{\centering\arraybackslash}p{\dimexpr.14\textwidth-2\tabcolsep}
      >{\centering\arraybackslash}p{\dimexpr.14\textwidth-2\tabcolsep}
    }
    Input & \hspace{1.2em}$2\times$ & \hspace{1.1em}$4\times$ & \hspace{.8em}$8\times$ & \hspace{.8em}$16\times$ & \hspace{.8em}$32\times$ & \hspace{.8em}$64\times$ \\
    \end{tabular}

    \centering
    \rotatebox{90}{\hspace{1.7em}Bicubic}
    \begin{subfigure}{.135\textwidth}
        \includegraphics[width=\linewidth]{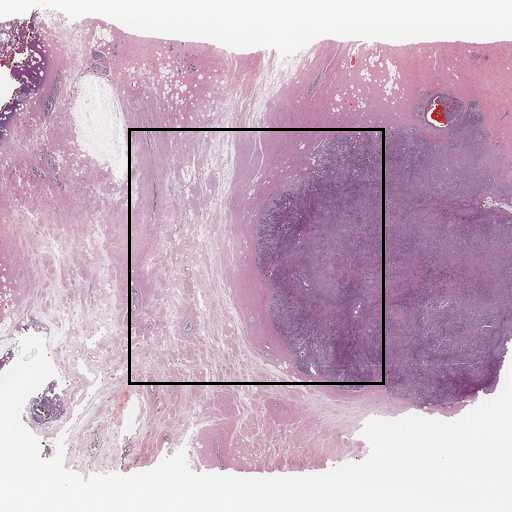}
    \end{subfigure}
    \hspace{0.01em}
    \begin{subfigure}{.135\textwidth}
        \includegraphics[width=\linewidth]{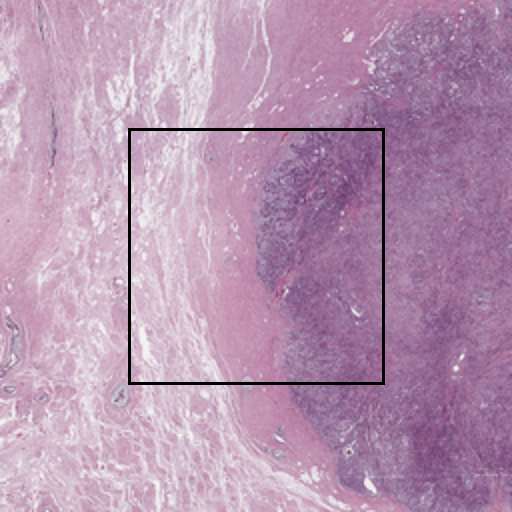}
    \end{subfigure}
    \begin{subfigure}{.135\textwidth}
        \includegraphics[width=\linewidth]{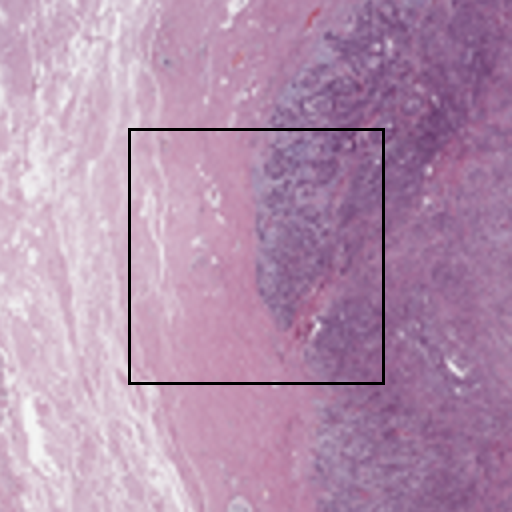}
    \end{subfigure}
    \begin{subfigure}{.135\textwidth}
        \includegraphics[width=\linewidth]{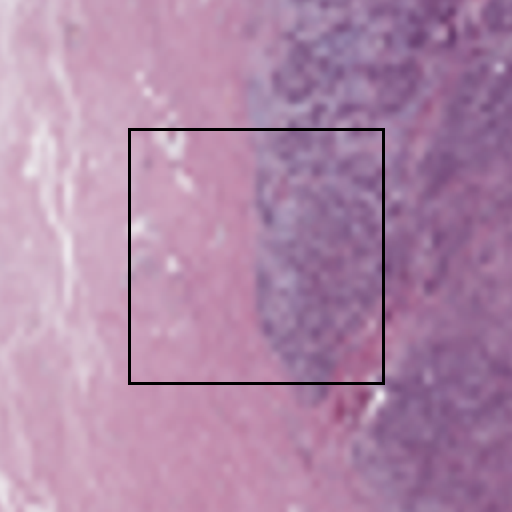}
    \end{subfigure}
    \begin{subfigure}{.135\textwidth}
        \includegraphics[width=\linewidth]{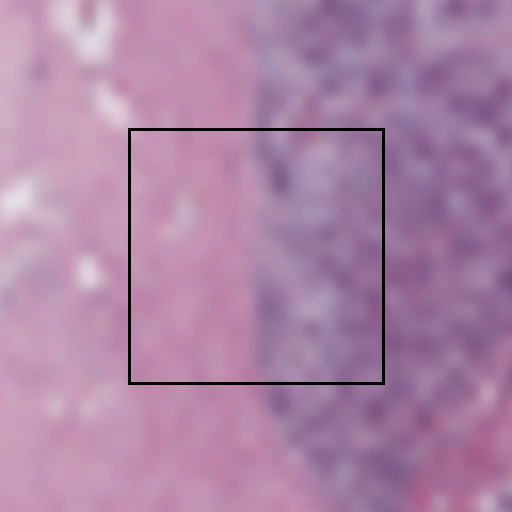}
    \end{subfigure}
    \begin{subfigure}{.135\textwidth}
        \includegraphics[width=\linewidth]{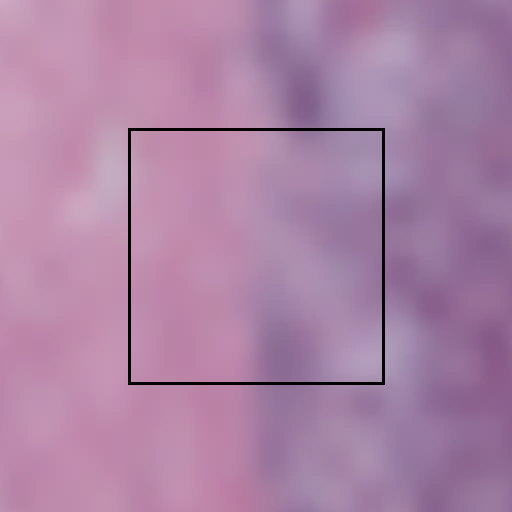}
    \end{subfigure}
    \begin{subfigure}{.135\textwidth}
        \includegraphics[width=\linewidth]{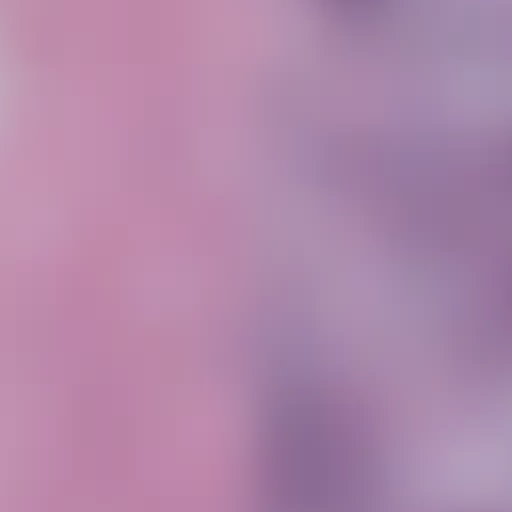}
    \end{subfigure}

    \rotatebox{90}{\hspace{1.9em}TV-L1}
    \begin{subfigure}{.135\textwidth}
        \includegraphics[width=\linewidth]{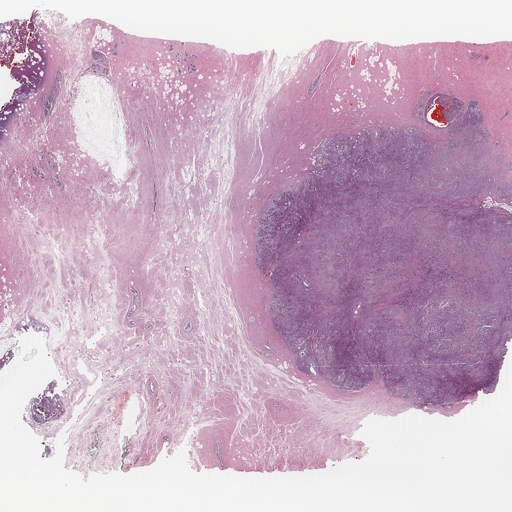}
    \end{subfigure}
    \hspace{0.01em}
    \begin{subfigure}{.135\textwidth}
        \includegraphics[width=\linewidth]{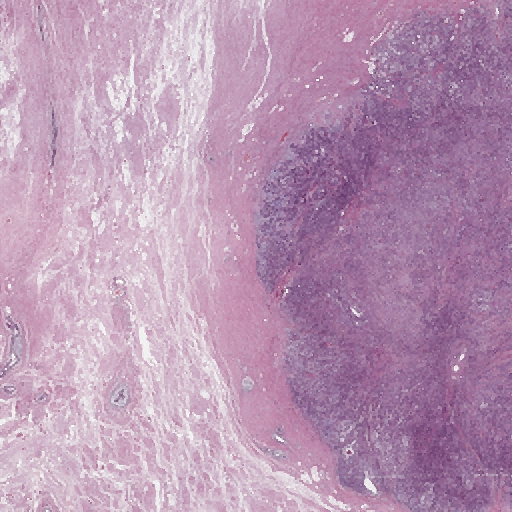}
    \end{subfigure}
    \begin{subfigure}{.135\textwidth}
        \includegraphics[width=\linewidth]{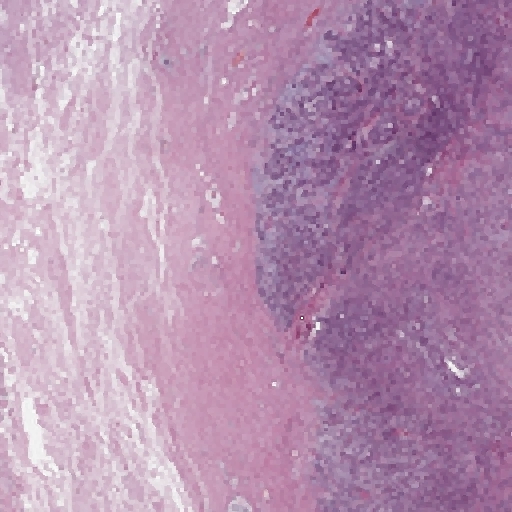}
    \end{subfigure}
    \begin{subfigure}{.135\textwidth}
        \includegraphics[width=\linewidth]{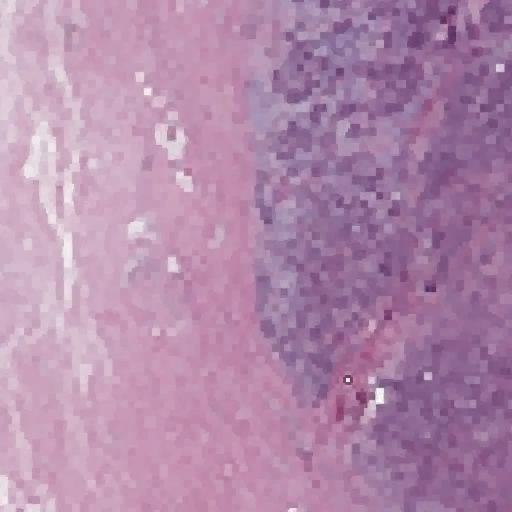}
    \end{subfigure}
    \begin{subfigure}{.135\textwidth}
        \includegraphics[width=\linewidth]{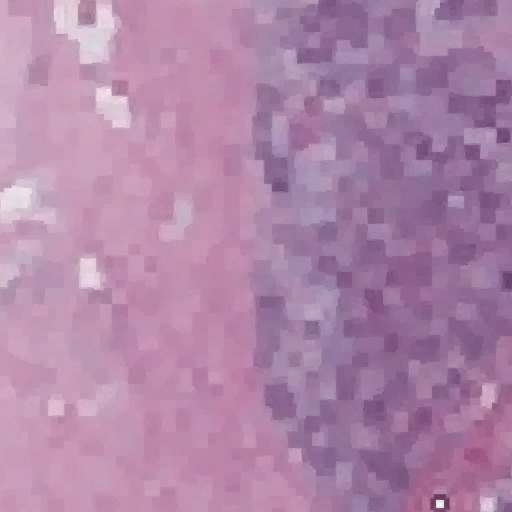}
    \end{subfigure}
    \begin{subfigure}{.135\textwidth}
        \includegraphics[width=\linewidth]{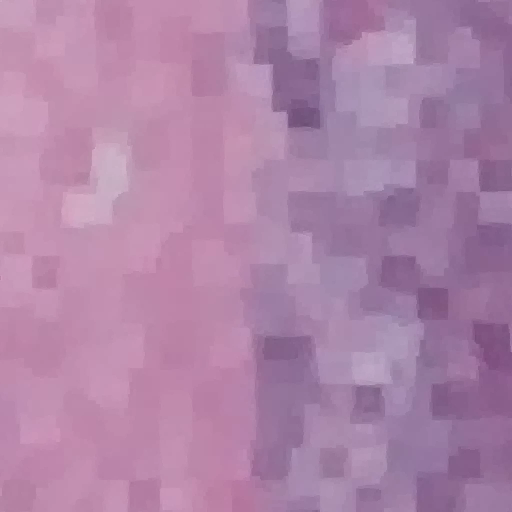}
    \end{subfigure}
    \begin{subfigure}{.135\textwidth}
        \includegraphics[width=\linewidth]{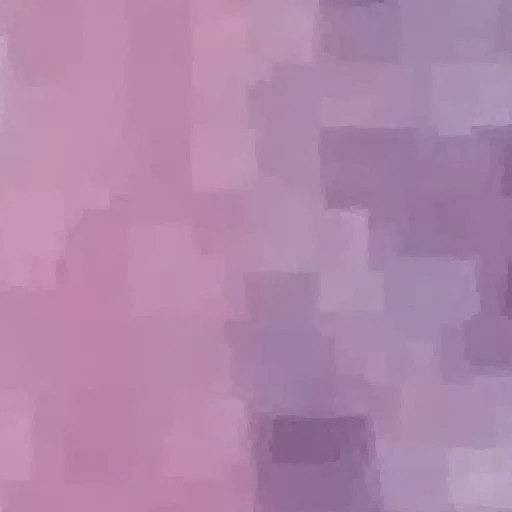}
    \end{subfigure}

    \rotatebox{90}{\hspace{1.9em}EDSR}
    \begin{subfigure}{.135\textwidth}
        \includegraphics[width=\linewidth]{fig/comparison_input/diff_model/slide.png}
    \end{subfigure}
    \hspace{0.01em}
    \begin{subfigure}{.135\textwidth}
        \includegraphics[width=\linewidth]{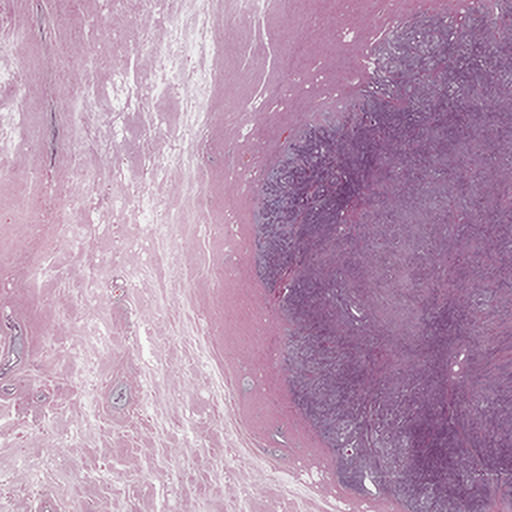}
    \end{subfigure}
    \begin{subfigure}{.135\textwidth}
        \includegraphics[width=\linewidth]{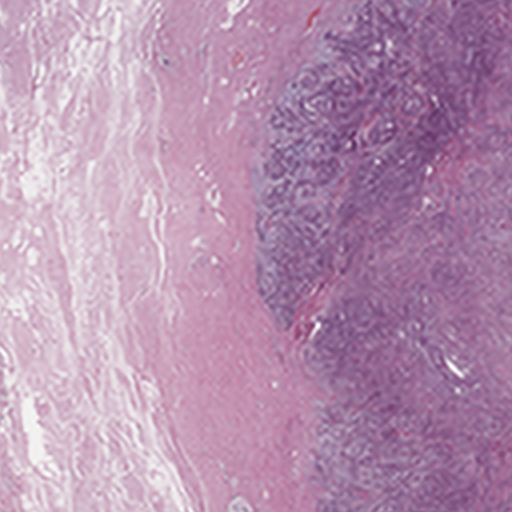}
    \end{subfigure}
    \begin{subfigure}{.135\textwidth}
        \includegraphics[width=\linewidth]{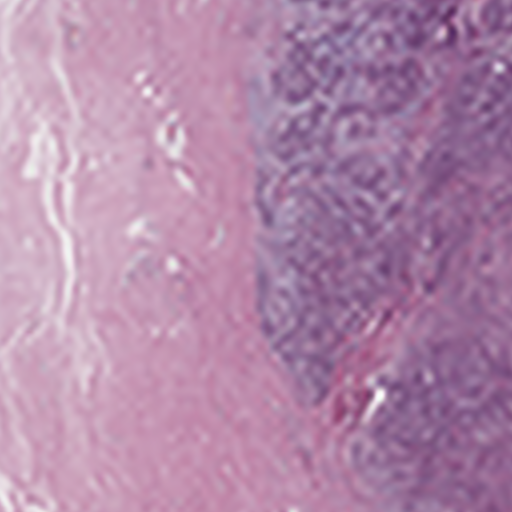}
    \end{subfigure}
    \begin{subfigure}{.135\textwidth}
        \includegraphics[width=\linewidth]{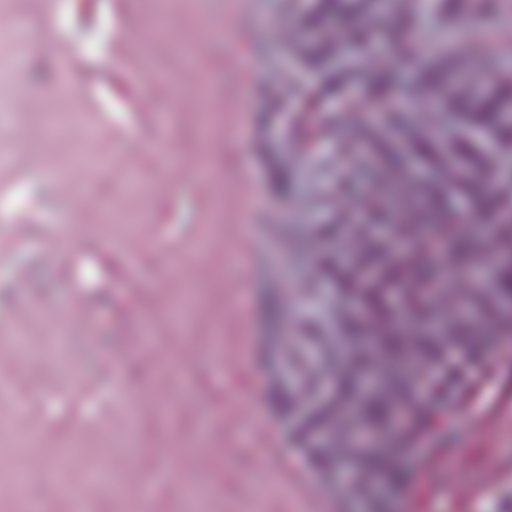}
    \end{subfigure}
    \begin{subfigure}{.135\textwidth}
        \includegraphics[width=\linewidth]{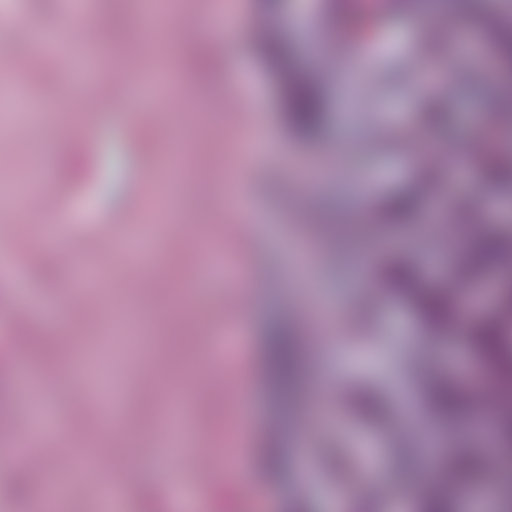}
    \end{subfigure}
    \begin{subfigure}{.135\textwidth}
        \includegraphics[width=\linewidth]{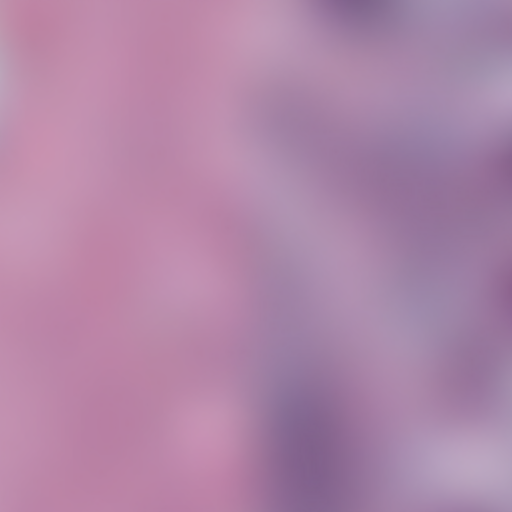}
    \end{subfigure}

    \rotatebox{90}{\hspace{1.05em}Ours, $r=0$}
    \begin{subfigure}{.135\textwidth}
        \includegraphics[width=\linewidth]{fig/comparison_input/diff_model/slide.png}
    \end{subfigure}
    \hspace{0.01em}
    \begin{subfigure}{.135\textwidth}
        \includegraphics[width=\linewidth]{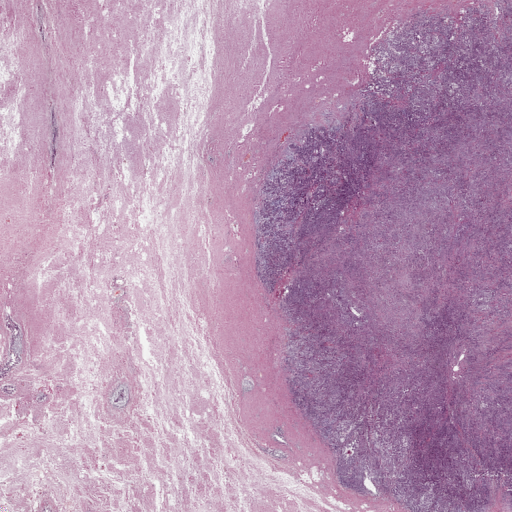}
    \end{subfigure}
    \begin{subfigure}{.135\textwidth}
        \includegraphics[width=\linewidth]{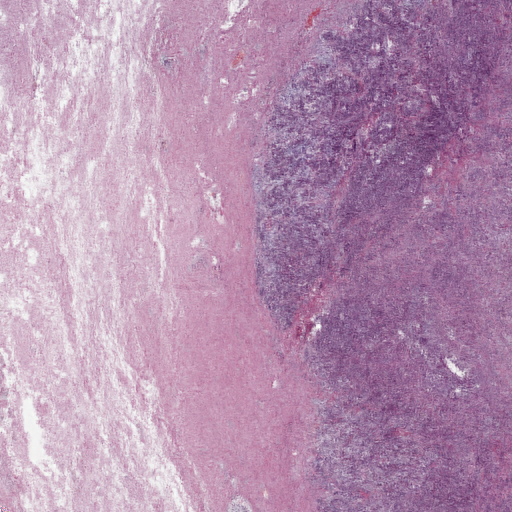}
    \end{subfigure}
    \begin{subfigure}{.135\textwidth}
        \includegraphics[width=\linewidth]{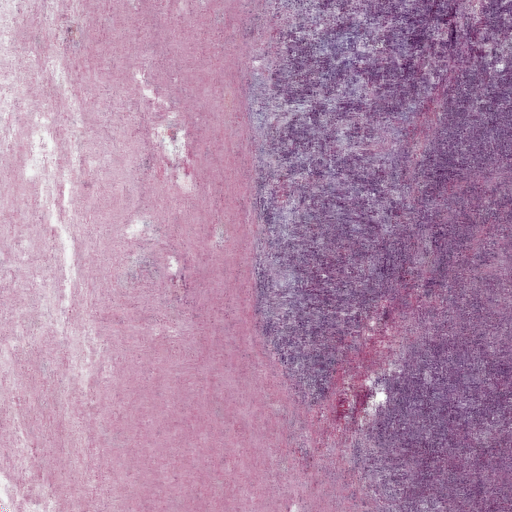}
    \end{subfigure}
    \begin{subfigure}{.135\textwidth}
        \includegraphics[width=\linewidth]{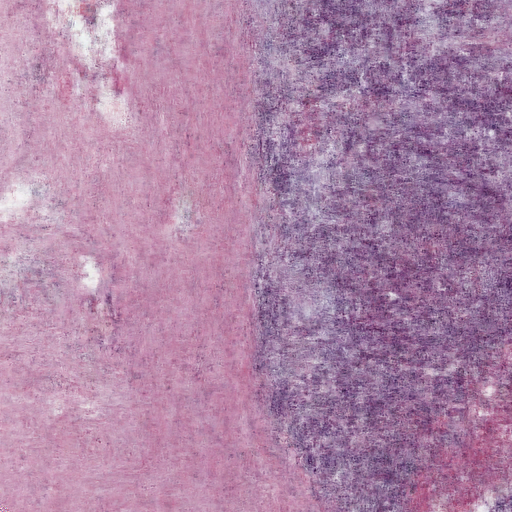}
    \end{subfigure}
    \begin{subfigure}{.135\textwidth}
        \includegraphics[width=\linewidth]{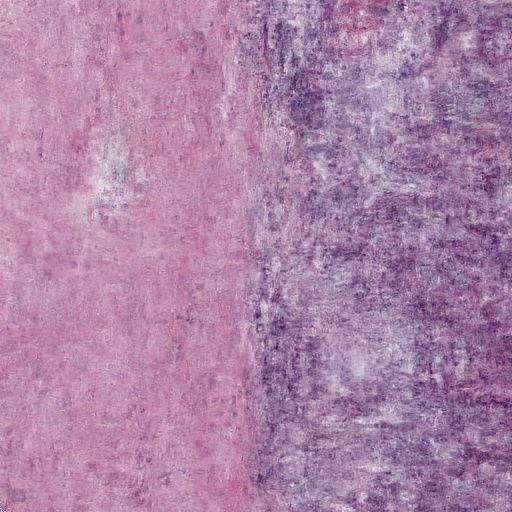}
    \end{subfigure}
    \begin{subfigure}{.135\textwidth}
        \includegraphics[width=\linewidth]{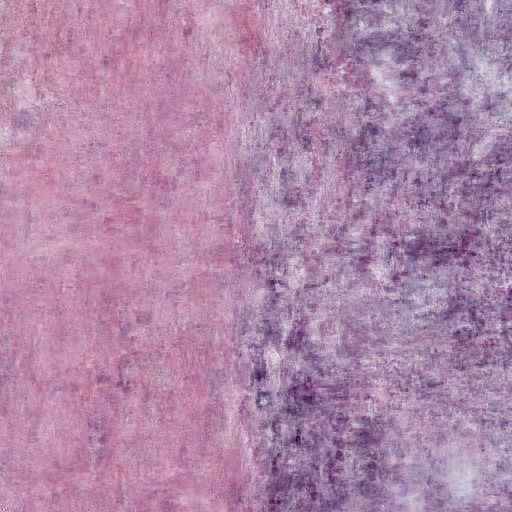}
    \end{subfigure}

    \rotatebox{90}{\hspace{.7em}Ours, $r=28$}
    \begin{subfigure}{.135\textwidth}
        \includegraphics[width=\linewidth]{fig/comparison_input/diff_model/slide.png}
    \end{subfigure}
    \hspace{0.01em}
    \begin{subfigure}{.135\textwidth}
        \includegraphics[width=\linewidth]{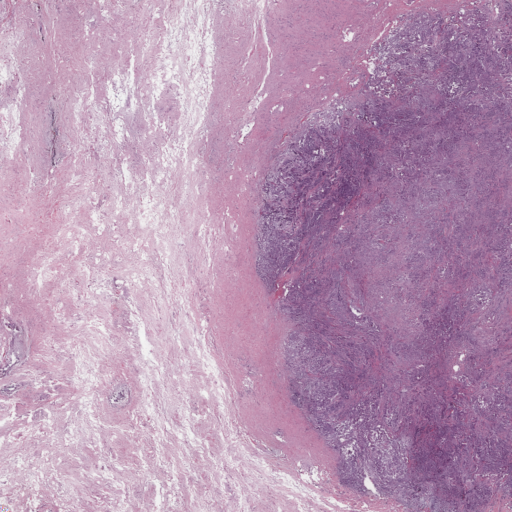}
    \end{subfigure}
    \begin{subfigure}{.135\textwidth}
        \includegraphics[width=\linewidth]{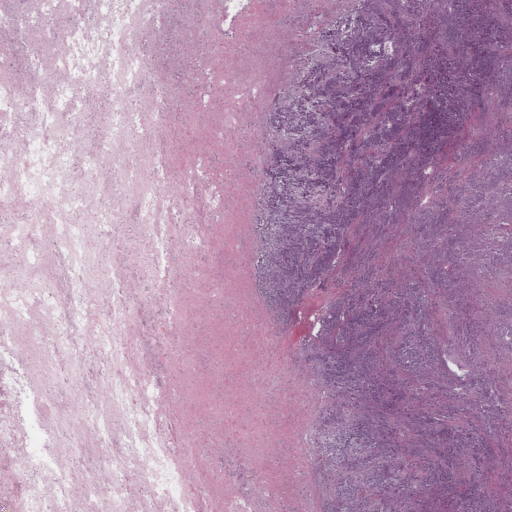}
    \end{subfigure}
    \begin{subfigure}{.135\textwidth}
        \includegraphics[width=\linewidth]{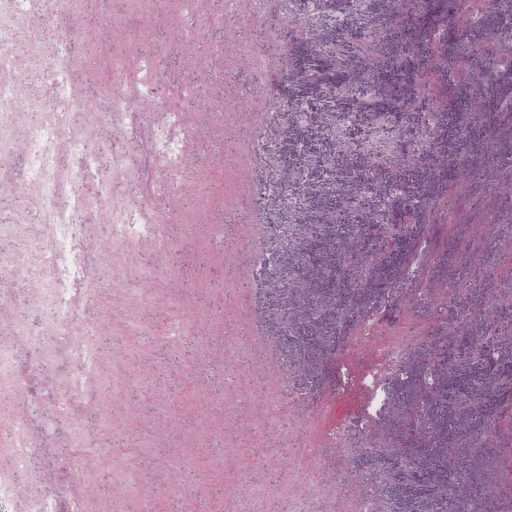}
    \end{subfigure}
    \begin{subfigure}{.135\textwidth}
        \includegraphics[width=\linewidth]{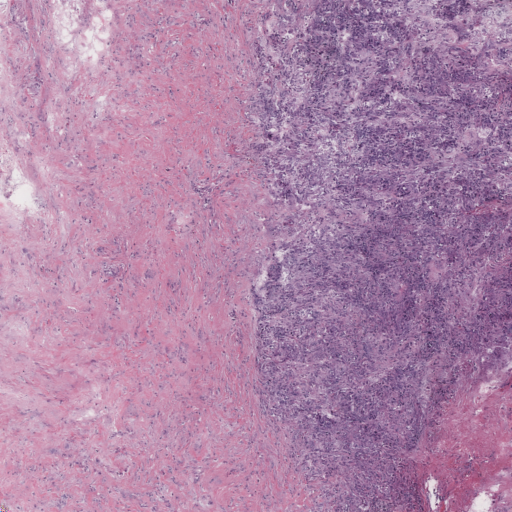}
    \end{subfigure}
    \begin{subfigure}{.135\textwidth}
        \includegraphics[width=\linewidth]{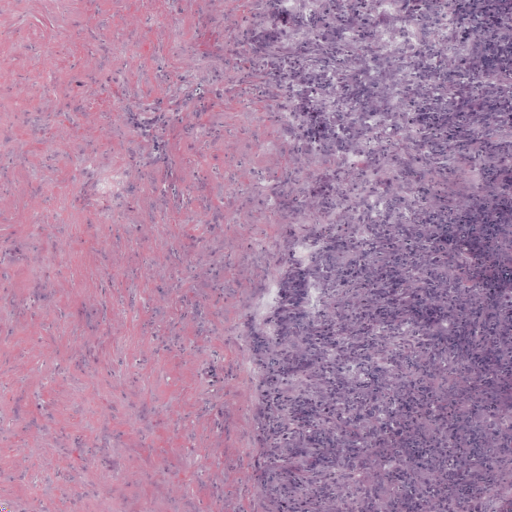}
    \end{subfigure}
    \begin{subfigure}{.135\textwidth}
        \includegraphics[width=\linewidth]{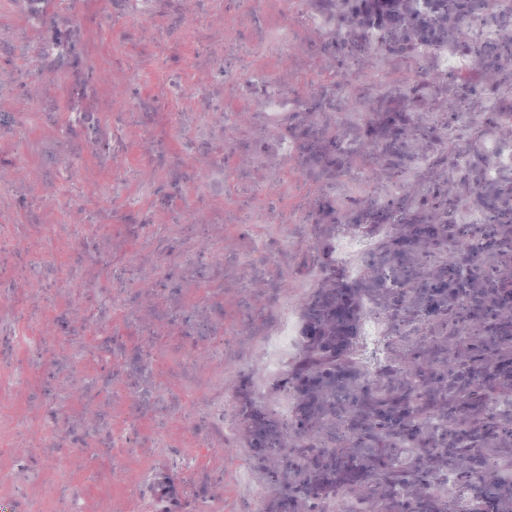}
    \end{subfigure}

    \caption{Comparison of our method with multiple super-resolution methods.
    The first column shows the input image, each subsequent column shows the upscaling result of a patch extracted from the center of the previous column.
    Best viewed digitally.}
    \label{fig:comparison}
\end{figure*}

  \section{Derivation of the super-resolution constraint}
    \label{sec:ss_comp}
  In the following, we show how to solve the optimization problem given in \cref{eq:optimization} using the method of Lagrangian multipliers.
  We begin with the problem
\begin{equation}
    \bar{\mathbf{u}} = \argmin_{\bar{\mathbf{u}}} \frac{1}{2}\|\mathbf{u}-\bar{\mathbf{u}}\|^2 \quad \text { s.t. } \mathbf{A}\bar{\mathbf{u}} = \mathbf{y}.
  \end{equation}
  To solve it, we introduce the Lagrangian
  \begin{equation}
    \mathcal{L}(\bar{\mathbf{u}}, \mathbf{\lambda}) = \frac{1}{2}\|\mathbf{u}-\bar{\mathbf{u}}\|^2 + \mathbf{\lambda}^T (\mathbf{A}\bar{\mathbf{u}} - \mathbf{y}),
    \end{equation}
    with lagrange multipliers $\lambda$.
    The gradient of the lagrangian is given as
  \begin{equation}
        \nabla_{\bar{\mathbf{u}}} \mathcal{L} =  \bar{\mathbf{u}} - \mathbf{u} + \mathbf{A}^T \mathbf{\lambda}.
    \end{equation}
    Furthermore, setting the gradient to zero, and solving for $\bar{\mathbf{u}}$ results in
  \begin{equation}
    \bar{\mathbf{u}} = \mathbf{u} - \mathbf{A}^T \lambda.
    \label{eq:lag}
  \end{equation}
  Inserting $\bar{\mathbf{u}}$ from \cref{eq:lag} into the constraint gives
  \begin{equation}
    \mathbf{y} = \mathbf{A} (\mathbf{u} - \mathbf{A}^T \lambda).
        \label{eq:solve}
  \end{equation}
  By solving \cref{eq:solve} for $\lambda$ we obtain
  \begin{equation}
    \lambda = (\mathbf{A} \mathbf{A}^T)^{-1}(\mathbf{A}\mathbf{u} - \mathbf{y}).
    \label{eq:lammm}
  \end{equation}
  And inserting $\lambda$ from \cref{eq:lammm} into \cref{eq:lag}, gives us a solution for $\bar{\mathbf{u}}$
  \begin{equation}
    \bar{\mathbf{u}} = \mathbf{u} - \mathbf{A}^{T}\left(\mathbf{A} \mathbf{A}^{T}\right)^{-1}(\mathbf{A} \mathbf{u}-\mathbf{y}),
  \end{equation}
  which can be simplified as
  \begin{equation}
    \begin{aligned}
     \bar{\mathbf{u}} &= \left(\mathbf{I} - \mathbf{A}^T(\mathbf{A} \mathbf{A}^T)^{-1} \mathbf{A} \right) \mathbf{u} + \mathbf{A}^T (\mathbf{A} \mathbf{A}^T)^{-1} \mathbf{y} \\
     &= \mathbf{u} - \mathbf{A}^{T} \left(\mathbf{A} \mathbf{A}^{T}\right)^{-1} \mathbf{A} \mathbf{u} - \mathbf{A}^{T}\left(\mathbf{A} \mathbf{A}^{T}\right)^{-1}{\mathbf{y}}.
    \end{aligned}
     \label{eq:above}
  \end{equation}
 Using the definition of the pseudoinverse $\mathbf{A}^{\dagger}$ for full row rank matrices
  \begin{equation}
    \mathbf{A}^{\dagger} = \mathbf{A}^T(\mathbf{A} \mathbf{A}^T)^{-1},
  \end{equation}
  we can further simplify \cref{eq:above} leading to our final solution
  \begin{equation}
    \begin{aligned}
     \bar{\mathbf{u}} &= (\mathbf{I} - \mathbf{A}^\dagger \mathbf{A}) \mathbf{u} + \mathbf{A}^{\dagger} \mathbf{y}.
    \end{aligned}
  \end{equation}

  \section{Scaling functions}\label{app:scale}
  In the following, we provide the full expressions of the noise level parametrized scaling functions in our diffusion model.
  Particularly, in our denoiser function \cref{eq:denoiser_skipped}
\begin{equation}
    D_\theta(\mathbf{x} ; \sigma, s) = c_{\text {skip }}(\sigma) \, \mathbf{x}+c_{\text {out }}(\sigma) \, F_\theta\big(c_{\text {in }}(\sigma) \, \mathbf{x} ; \sigma, s \big),
\end{equation}
and our loss \cref{eq:loss}
\begin{equation}
    \mathbb{E}_{s, \tilde{\mathbf{x}}, \sigma, \mathbf{n}}\big[\lambda(\sigma) \|D_{\theta}(\tilde{\mathbf{x}} + \mathbf{n} ; \sigma, s) - \tilde{\mathbf{x}} \|_2^2 \big],
\end{equation}
we set
\begin{equation}
    c_{\text{skip}}(\sigma) = \sigma_{\text {data }}^2 /\left(\sigma^2+\sigma_{\text {data }}^2\right),
\end{equation}
\begin{equation}
    c_{\text{out}}(\sigma) = \sigma \cdot \sigma_{\text {data }} / \sqrt{\sigma_{\text {data }}^2+\sigma^2},
\end{equation}
\begin{equation}
    c_{\text{in}}(\sigma) = 1 / \sqrt{\sigma^2+\sigma_{\text {data }}^2},
\end{equation}
and
\begin{equation}
    \lambda(\sigma) = \sigma^{-2}+\frac{1}{\sigma_{\text {data }}^2},
\end{equation}
where $\sigma_{\text{data}}$ is the standard deviation of our training data.
We set $\sigma_{\text{data}}=0.5$, which is simply done through the normalization of training images.
A detailed discussion and derivations of these noise level parametrized scaling functions are provided by Karras \etal \cite{karras2022elucidating}.
In essence, the input scaling $c_{\text{in}}(\sigma)$ is set such that the inputs of $F_{\theta}$ have unit variance.
         The output scaling $c_{\text{out}}(\sigma)$ is set such that the effective training target of $F_{\theta}$ has unit variance.
         The skip-connection scaling $c_{\text {skip }}(\sigma)$ is set such that the errors of $F_{\theta}$ are amplified as little as possible.
         And the loss weighting $\lambda(\sigma)$ weighs loss terms equally across all noise levels $\sigma$.

   \section{Data preprocessing}
     When sampling patches from \glspl{WSI}, we only consider patches covering at least $10 \%$ tissue area.
     To segment tissue from the background, we use FESI \cite{bug2015foreground}.

\section{Comparison with different super-resolution approaches}
In this section, we compare upscaling with our approach to established super-resolution methods.
To this end, we apply multiple iterations of $2\times$ upscaling on an initial ${512\!\times\!512}$-sized image.
In each iteration, we upscale a ${512\!\times\!512}$ patch extracted from the centre of the previous iteration's output.
We compare with super-resolution approaches that follow multiple paradigms:
TV-L1, which is not learning-based;
EDSR \cite{lim2017enhanced}, which is learning-based but not generative;
and our method, which is learning-based and generative.
For EDSR, we retrained the model using the same data as our method.
Additionally, as a baseline, we also show bicubic interpolation.

For our method, we show results with a relaxation parameter $r=0$ and with $r=28$.
As discussed in the main paper, without relaxation, \ie $r=0$, our method closely resembles the zero-shot super-resolution approach of DDNM \cite{wang2022zero}, where the super-resolution constraint has to be satisfied strictly.
Contrarily, with the relaxation parameter $r>0$, the model is not strictly bound to the super-resolution constraint, allowing for a trade-off between consistency with the low-resolution input image and introducing new details.

\Cref{fig:comparison} shows the results of our comparison.
TV-L1 super-resolution produces sharper results than bicubic interpolation but still gives unsatisfying results for larger magnifications.
Similarly, EDSR fails to produce reasonable results for larger magnifications.
The results of our method without relaxation are much sharper than TV-L1 and EDSR.
However, particularly at larger magnifications, the results no longer retain the structure of histopathological images.
Note how individual cells are barely visible at $64\times$ magnification.
In contrast, with relaxation, even at large magnifications, results resemble the structure of histopathological images much better, e.g. individual cells are clearly distinguishable.

\section{Sampling}
When sampling \glspl{WSI}, we segment the initial image $\mathbf{z}_0$ into tissue and background areas using FESI \cite{bug2015foreground}.
And then run the coarse-to-fine scheme only on patches that cover tissue area.
This helps us to reduce the overall sampling time by skipping areas containing background.
When stitching patches back together, we fill background patches with the background colour extracted from the segmentation.

   \section{Network}
   For the network $F_{\theta}(\mathbf{x}; \sigma, s)$, we used the U-Net backbone from the implementation of Karras \etal \cite{karras2022elucidating}, which is based on the network of DDPM++ \cite{song2020score}.
   \cref{tb:train} shows the parameters we used. 
   We did the additional conditioning with the spatial resolution $s$,
   in the same way as the noise conditioning $\sigma$ is implemented in the network.
   Hence, we compute a sinusoidal positional encoding of the spatial resolution $s$ in $\SI{}{\micro \metre} / \SI{}{\px}$ and push the result through embedding layers.
   We then simply add the spatial resolution embedding to the embedding of the noise and use the result for following computations instead of the plain noise embedding.
 
   \begin{table}[h]
    \centering
    \begin{tabular}{ll}
        \toprule
        \textbf{Parameter} & \textbf{Value} \\
        \midrule
        Channel multiplier & 64 \\
        Channel factor per resolution & 0.5-1-1-2-2-4-4 \\
        Residual blocks per resolution & 2 \\
        Attention resolutions & \{32-16-8\} \\
        Attention heads & 4 \\
        Dropout probability & 10\% \\
        \bottomrule
    \end{tabular}
    \caption{Network parameters}
    \end{table}

   \section{Training}
   \cref{tb:train} shows the parameters we used for training.
   Noise $\sigma$ during training was sampled from a log-normal distribution $\ln(\sigma)\sim \mathcal{N}(P_{\text{mean}}, P_{\text{std}}^2)$.
   \begin{table}[h]
    \centering
    \begin{tabular}{ll}
        \toprule
        \textbf{Parameter} & \textbf{Value} \\
        \midrule
        Learning rate & \(1 \times 10^{-4}\) \\
        Optimizer & Adam \\
        Batch size & 64 \\
        $\sigma_{\text{min}}$ & 0.002 \\
        $\sigma_{\text{max}}$ & 80 \\
        $\rho$ & 7 \\
        $P_{\text{std}}$ & -1.2 \\
        $P_{\text{max}}$ & 1.2 \\
        \bottomrule
    \end{tabular}
    \caption{Training hyperparameters}
    \label{tb:train}
    \end{table}

\section{Downscaling operator}
In the following, we show how the average-pooling operator $\mathbf{A}$ and its pseudoinverse $\mathbf{A}^{\dagger}$ from \cref{eq:soll} can be implemented in PyTorch \cite{wang2022zero}.
    {
    \begin{lstlisting}
def PatchUpsample(x, scale):
    n,c,h,w = x.shape
    x = torch.zeros(n,c,h,scale,w,scale) + x.view(n,c,h,l,w,l)
    return x.view(n,c,scale*h,scale*w)

A = torch.nn.AdaptiveAvgPool2d(())
Ap = lambda z: PatchUpsample(z, scale)
    \end{lstlisting}
}

   \section{User study - additional discussion}
\begin{figure}[t]
        \begin{center}
            \includegraphics[width=.48\textwidth]{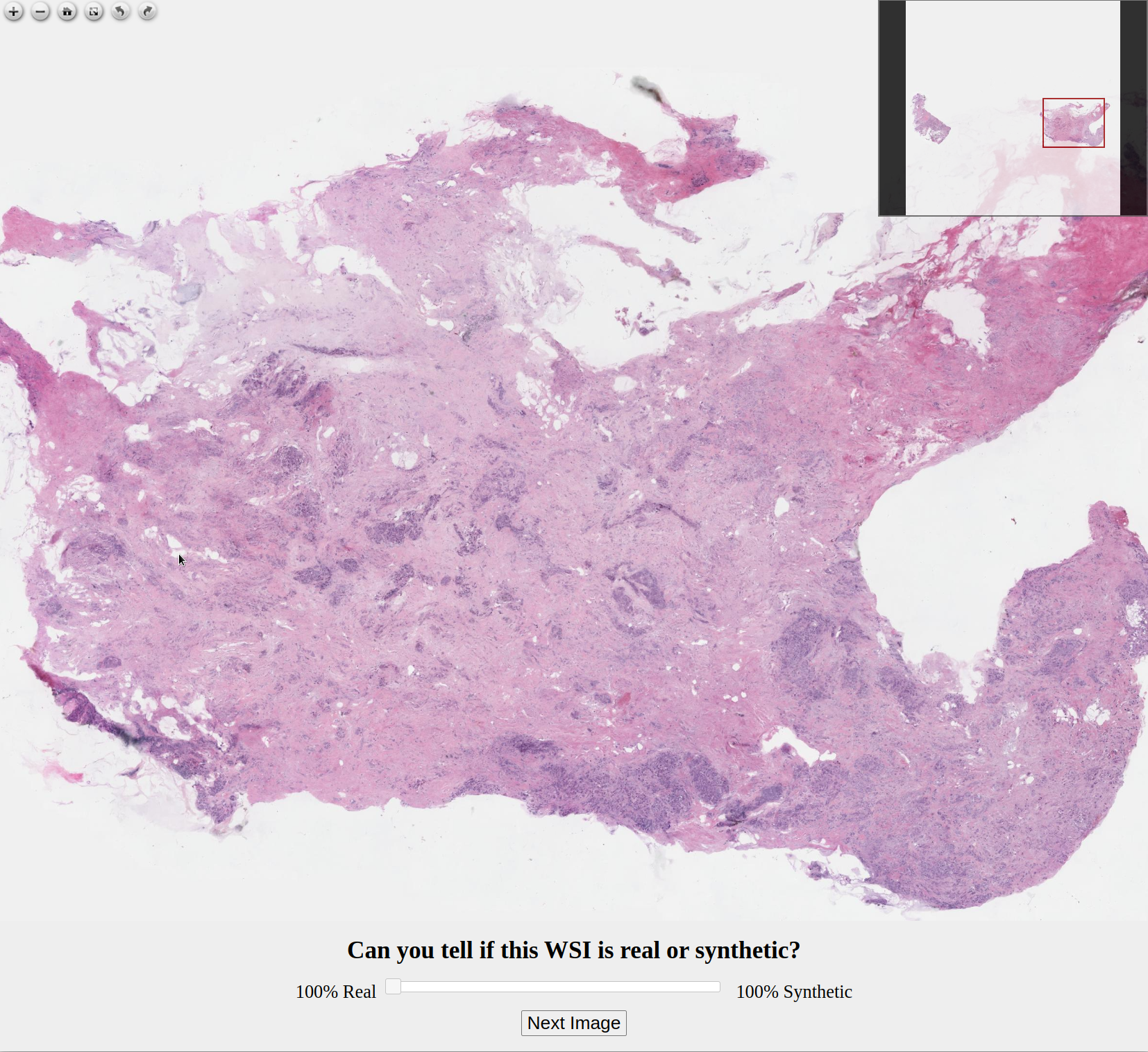}
        \end{center}
        \caption{Screenshot of our user study interface. The participants could freely navigate the shown \glspl{WSI} through their full magnification range.}
        \label{fig:user_study}
    \end{figure}

\begin{figure}[t]
        \begin{center}
            \includegraphics[width=.5\textwidth]{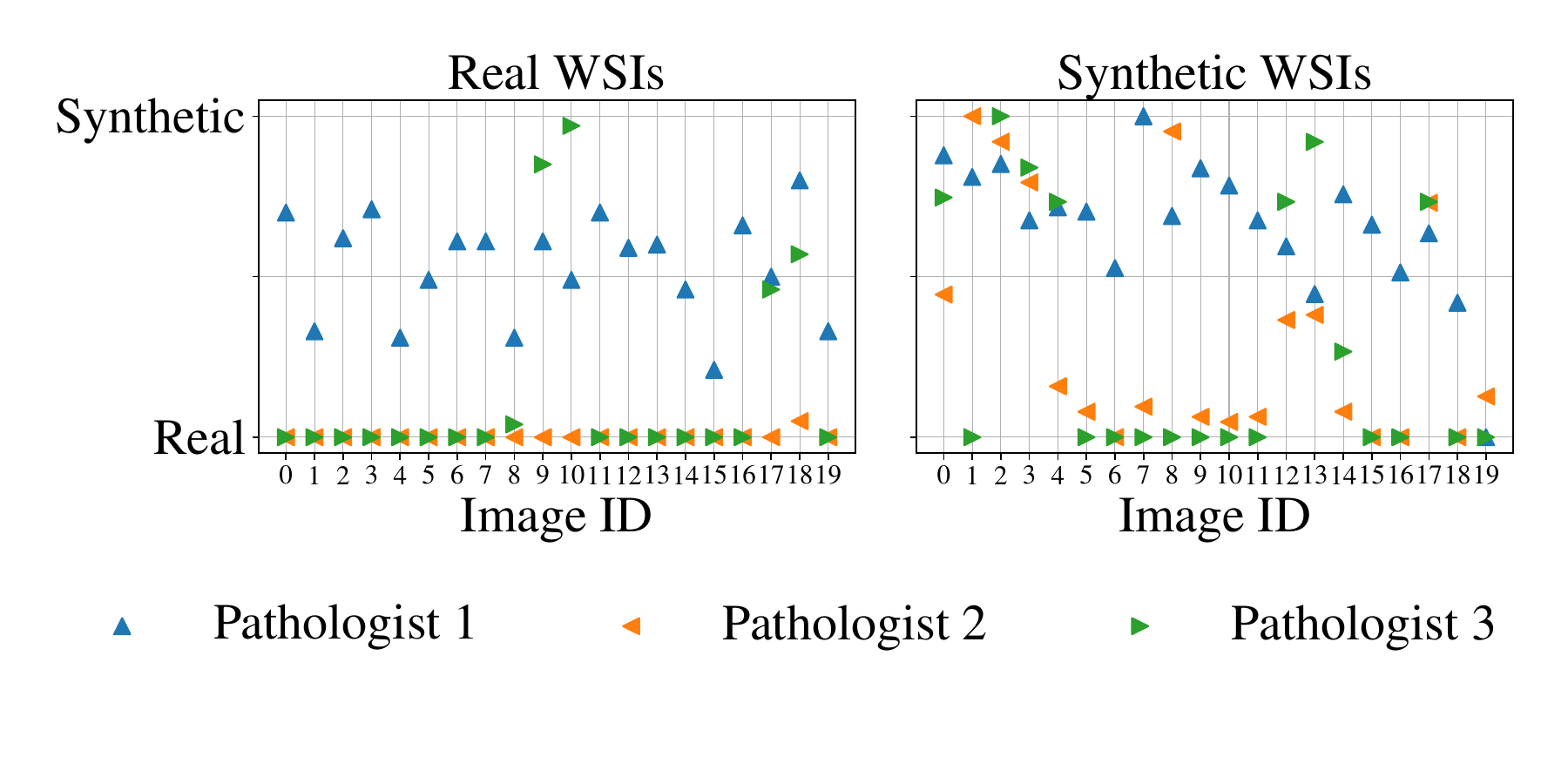}
        \end{center}
        \vspace{-2.0em}
        \caption{User study results with Image IDs. The IDs of the synthetic \glspl{WSI} correspond to the filenames in the provided download, and \cref{tab:tcga} maps the IDs of the real \glspl{WSI} to the respective IDs in the TCGA-BRCA dataset.}
        \label{fig:user_study_extended}
    \end{figure}

   \Cref{fig:user_study_extended} shows a visualization of the user study results with the respective IDs for each individual WSI.
   We provide a download to all 20 synthetic \glspl{WSI} of the user study\footnote{\label{note1}\url{https://drive.google.com/file/d/1VpNFGgcw2iEYY4cbHrskQsjjwHMy47A9/view?usp=sharing}}.
   To open the downloaded \glspl{WSI}, make sure to use an appropiate viewer, \eg QuPath\footnote{\url{https://qupath.github.io/}}.
   \Cref{tab:tcga} maps the IDs in the user study to the respective file IDs in the TCGA-BRCA dataset.
   Furthermore, \cref{fig:user_study} shows a screenshot of the interface we used for the study.

   In addition to the discussion in the main paper, we want to add a few remarks about the user study results. 
   Upon examining the results, one can see noticeable differences in the performance of the three pathologists when identifying the synthetic slides. 
   The first pathologist consistently gave ratings with a high degree of uncertainty. 
   In contrast, the other two seemed more confident in their decisions. 
   Notably, while the first pathologist correctly identified all the slides from the TCGA as real, the third pathologist mistakenly classified a few TCGA slides as synthetic with high certainty.
Even though there was a tendency for the pathologists to identify the synthetic slides, this suggests that it was not trivial for the pathologists to differentiate the images. 
Therefore, we conclude that most synthetic WSIs did not contain major, prominent image artefacts. 
This suggests that grid-shift was effective at preventing stitching artifacts and that our diffusion model did not generate completely pathologically unplausible structures.

    \begin{table}[th]
        \tiny
        \centering
        \begin{tabular}{ll}
            \toprule
            \textbf{User study ID} & \textbf{TCGA-BRCA ID} \\
            \midrule
            \texttt{0} & \texttt{TCGA-A7-A4SD-11A-03-TS3.3781BE68-0CC3-446C-9DA9-35EC6FA954E4} \\
            \texttt{1} & \texttt{TCGA-A7-A6VX-01Z-00-DX1.F74DA243-C65A-4997-BCA0-F1C89675978C} \\
            \texttt{2} & \texttt{TCGA-A8-A09I-01A-02-BS2.ca9aacf2-573b-4af2-bc50-5213526eb3a3}\\
            \texttt{3} & \texttt{TCGA-AN-A0FS-01A-01-TSA.ec030e02-fd7d-4683-803d-830ee80d8173} \\
            \texttt{4} & \texttt{TCGA-AO-A03U-01B-02-BSB.dcb167f4-c3ab-4dcc-8f40-41c4ce453847} \\
            \texttt{5} & \texttt{TCGA-AO-A0J5-01Z-00-DX1.20C14D0C-1A74-4FE9-A5E6-BDDCB8DE7714} \\
            \texttt{6} & \texttt{TCGA-AR-A0TR-01Z-00-DX1.BBCE653F-7DD0-4830-BAD3-C06207A93853} \\
            \texttt{7} & \texttt{TCGA-B6-A0IM-01A-01-BSA.e4fce1ac-0800-4e45-a3bc-f9bcb2ea825f} \\
            \texttt{8} & \texttt{TCGA-B6-A1KC-01Z-00-DX1.4DD3E48B-F434-499F-9FF1-0FFD2883A375} \\
            \texttt{9} & \texttt{TCGA-BH-A0BF-11A-02-TSB.6e4bf881-a29f-4fb4-b38c-5bebe44368ec} \\
            \texttt{10} & \texttt{TCGA-BH-A0DD-11A-01-BSA.e9aae98d-ecf8-4d48-b1ca-f349013f2c42} \\
            \texttt{11} & \texttt{TCGA-C8-A27B-01Z-00-DX1.5A8A14E8-6430-4147-9C71-805024E098CB} \\
            \texttt{12} & \texttt{TCGA-C8-A8HP-01A-01-TSA.C1048607-5CC7-4798-AA55-55C78B31C10D} \\
            \texttt{13} & \texttt{TCGA-E2-A15H-01A-01-TSA.6ba57309-1e15-4a84-98ad-5e8f02688a96} \\
            \texttt{14} & \texttt{TCGA-E2-A15M-01A-01-TSA.41d14b10-8567-4f43-a5a8-b952d859c70f} \\
            \texttt{15} & \texttt{TCGA-E9-A229-01Z-00-DX1.5B448B88-DA0C-44FF-87B3-20649A4A26FE} \\
            \texttt{16} & \texttt{TCGA-EW-A1OX-01A-01-TSA.74283185-7c47-44ce-8904-1a121870104e} \\
            \texttt{17} & \texttt{TCGA-EW-A1P5-01A-01-TSA.0fdc58ed-1cbd-4f60-839e-c12e1450e431} \\
            \texttt{18} & \texttt{TCGA-GM-A2DI-01A-03-TSC.DB9E24D8-2B07-483E-A490-2B64240EFCEE} \\
            \texttt{19} & \texttt{TCGA-OL-A66O-01Z-00-DX1.5F1E4C60-5CE8-41B4-A94D-4AA80D9253F9} \\
            \bottomrule
        \end{tabular}
        \caption{Mapping between the \gls{WSI} IDs in the user study and their IDs in the TCGA-BRCA dataset.}
        \label{tab:tcga}
    \end{table}

   \section{Additional examples}
   In the following, we show additional \glspl{WSI} generated by our method.
   The shown patches are resized to $512\!\times\!512$.
   To get a full impression about the quality of the generated \glspl{WSI}, download the full-resolution \glspl{WSI} from the user study.

   \newpage
   \begin{figure}[h!]
    \centering
    \includegraphics[width=0.9\linewidth]{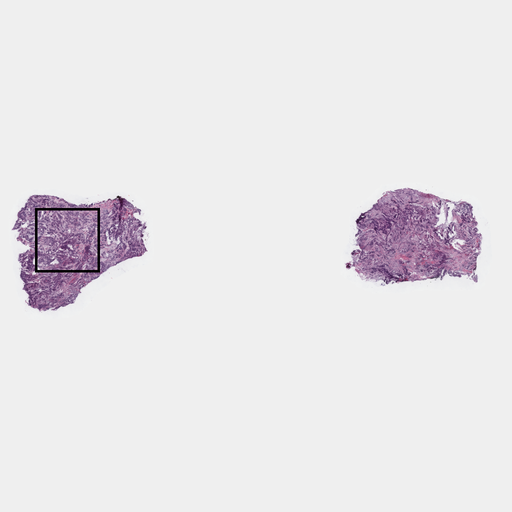}
    \vspace{0.1em}

    \includegraphics[width=0.9\linewidth]{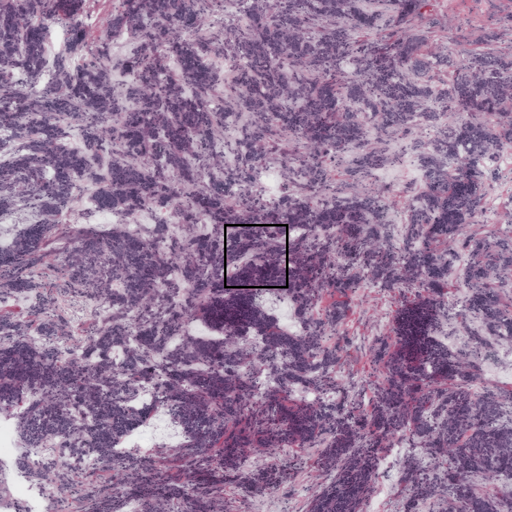}
    \vspace{0.1em}

    \includegraphics[width=0.9\linewidth]{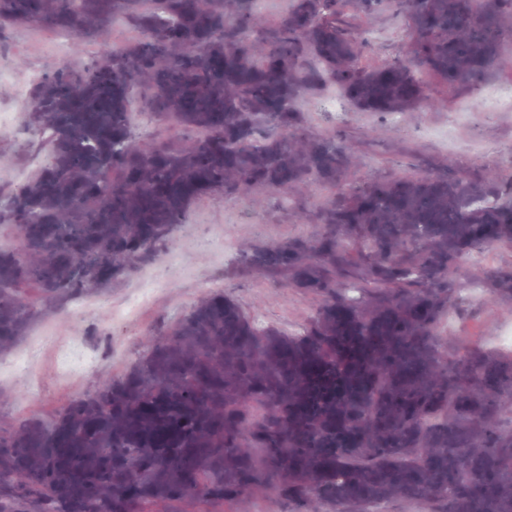}
    \vspace{-0.6em}
    \caption{Synthetic \gls{WSI}}
    \end{figure}

   \begin{figure}[h!]
    \centering
    \includegraphics[width=0.9\linewidth]{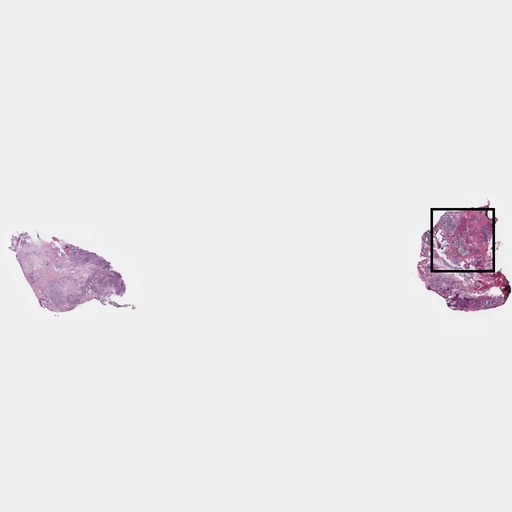}
    \vspace{0.1em}

    \includegraphics[width=0.9\linewidth]{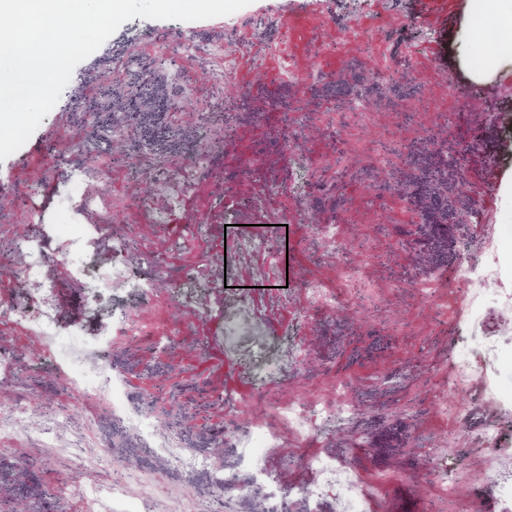}
    \vspace{0.1em}

    \includegraphics[width=0.9\linewidth]{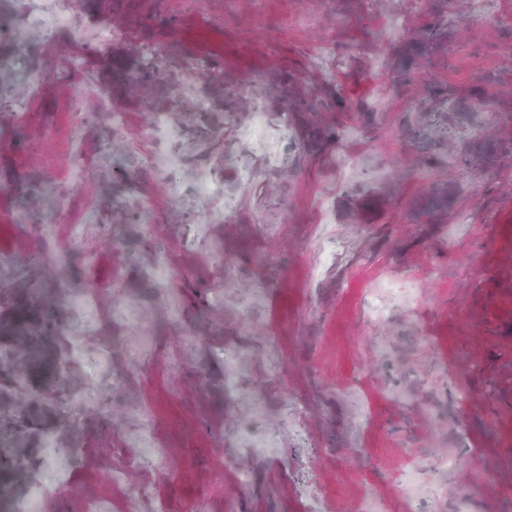}
    \vspace{-0.6em}
    \caption{Synthetic \gls{WSI}}
    \end{figure}

   \begin{figure}[h!]
    \centering
    \includegraphics[width=0.9\linewidth]{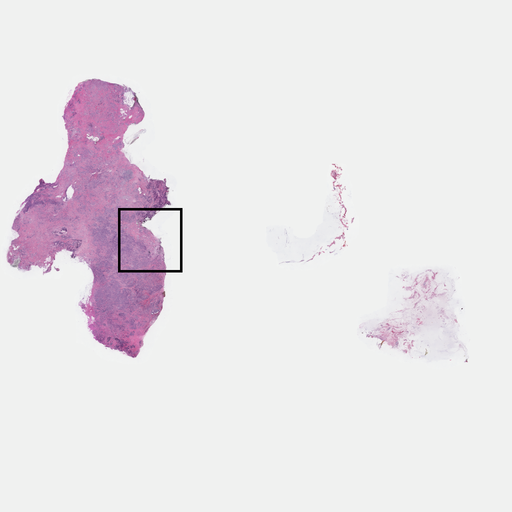}
    \vspace{0.1em}

    \includegraphics[width=0.9\linewidth]{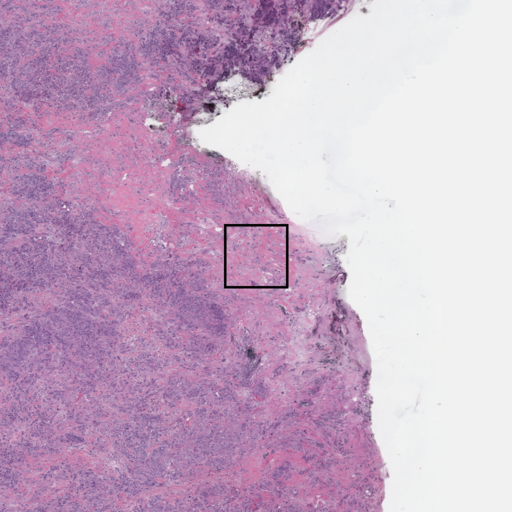}
    \vspace{0.1em}

    \includegraphics[width=0.9\linewidth]{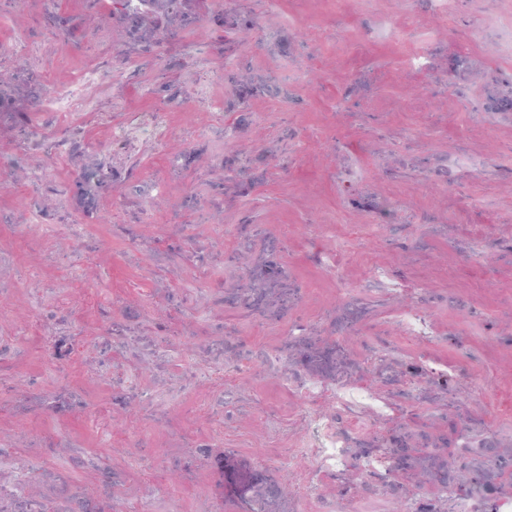}
    \vspace{-0.6em}
    \caption{Synthetic \gls{WSI}}
    \end{figure}

   \begin{figure}[h!]
    \centering
    \includegraphics[width=0.9\linewidth]{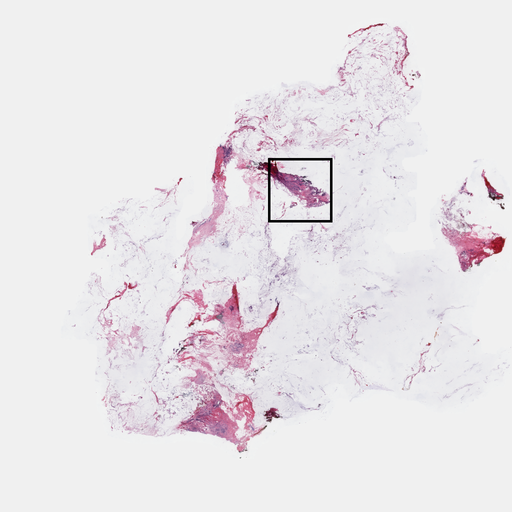}
    \vspace{0.1em}

    \includegraphics[width=0.9\linewidth]{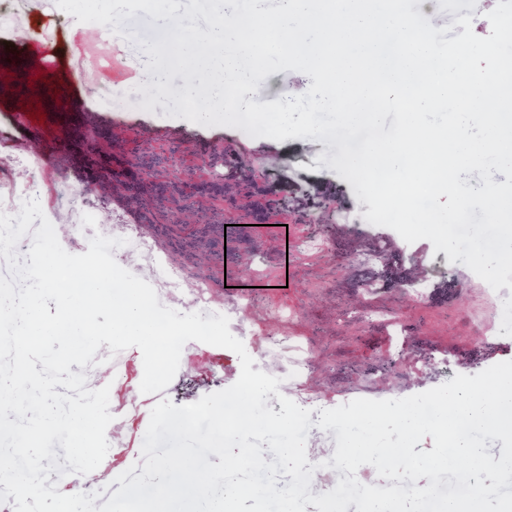}
    \vspace{0.1em}

    \includegraphics[width=0.9\linewidth]{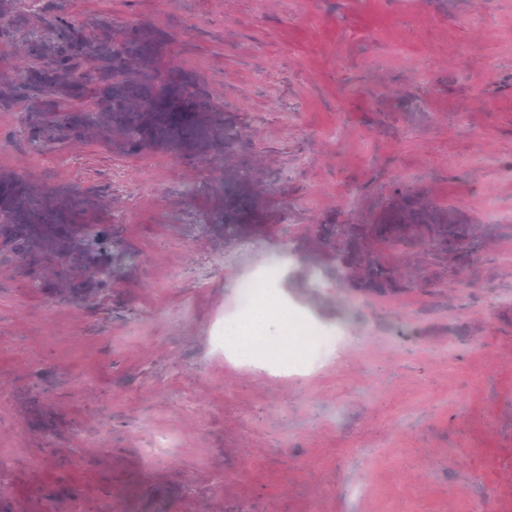}
    \vspace{-0.6em}
    \caption{Synthetic \gls{WSI}}
    \end{figure}

   \begin{figure}[h!]
    \centering
    \includegraphics[width=0.9\linewidth]{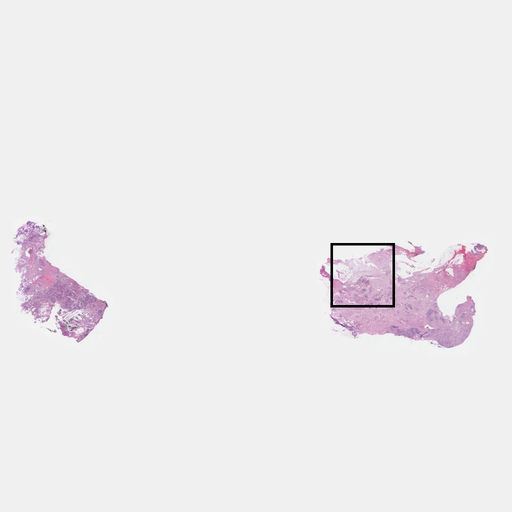}
    \vspace{0.1em}

    \includegraphics[width=0.9\linewidth]{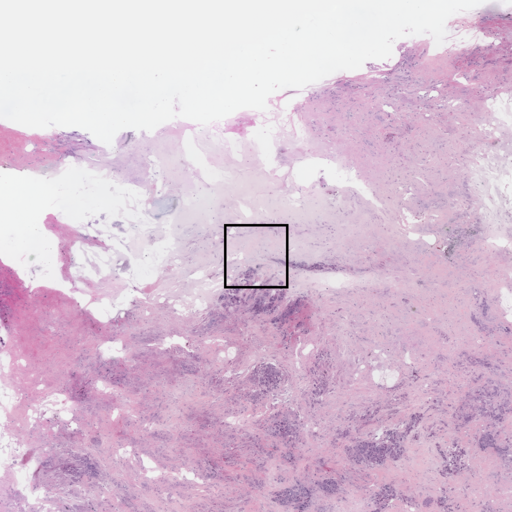}
    \vspace{0.1em}

    \includegraphics[width=0.9\linewidth]{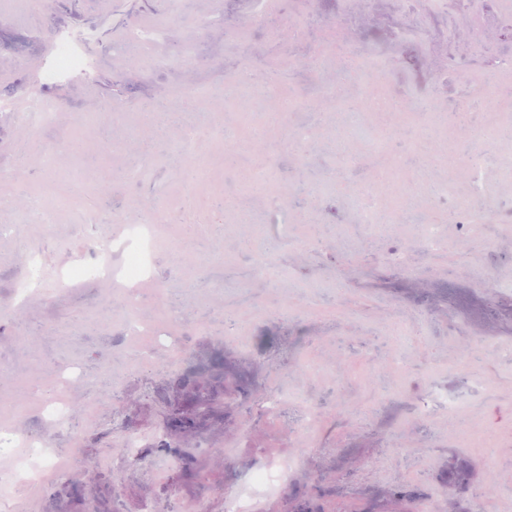}
    \vspace{-0.6em}
    \caption{Synthetic \gls{WSI}}
    \end{figure}

   \begin{figure}[h!]
    \centering
    \includegraphics[width=0.9\linewidth]{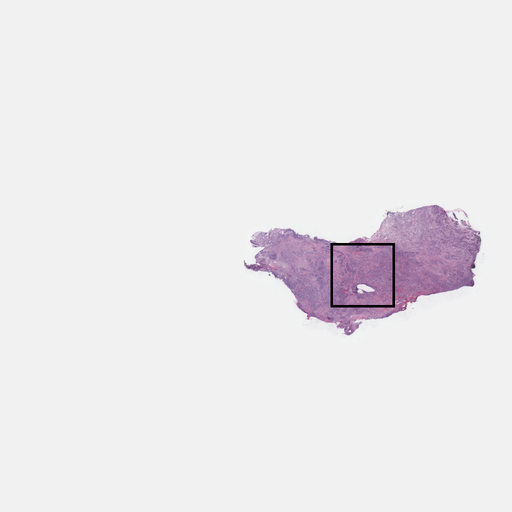}
    \vspace{0.1em}

    \includegraphics[width=0.9\linewidth]{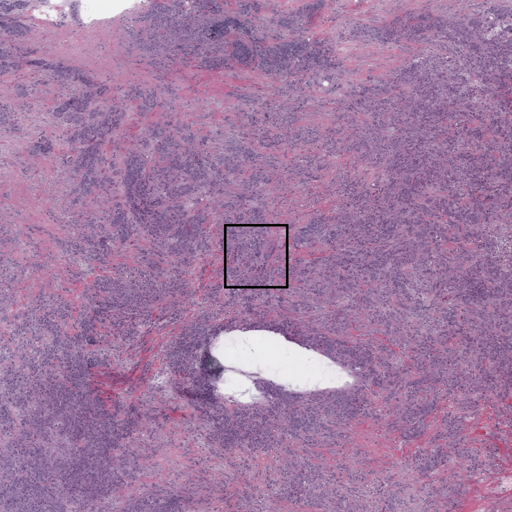}
    \vspace{0.1em}

    \includegraphics[width=0.9\linewidth]{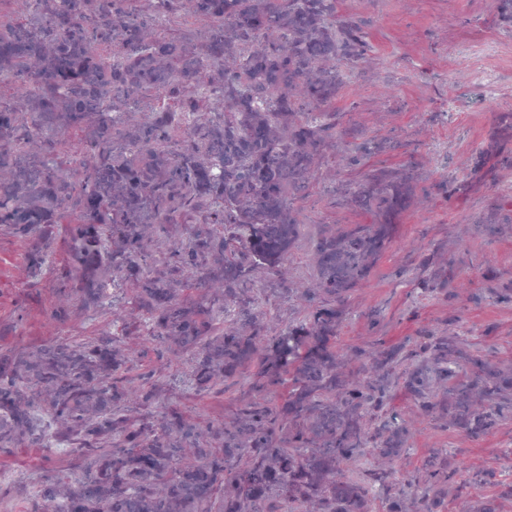}
    \vspace{-0.6em}
    \caption{Synthetic \gls{WSI}}
    \end{figure}

   \begin{figure}[h!]
    \centering
    \includegraphics[width=0.9\linewidth]{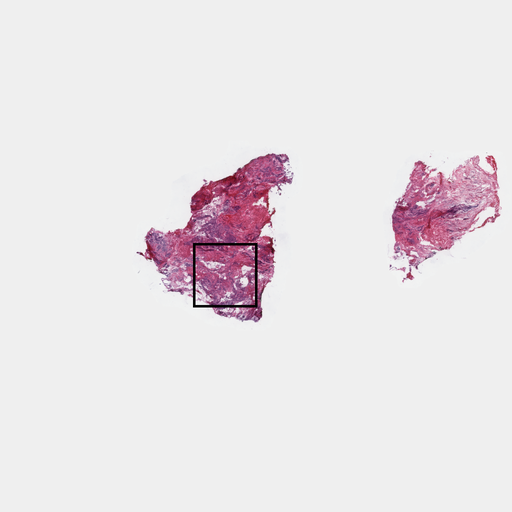}
    \vspace{0.1em}

    \includegraphics[width=0.9\linewidth]{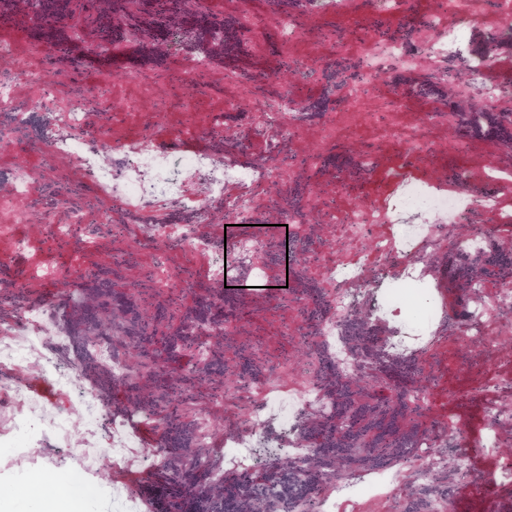}
    \vspace{0.1em}

    \includegraphics[width=0.9\linewidth]{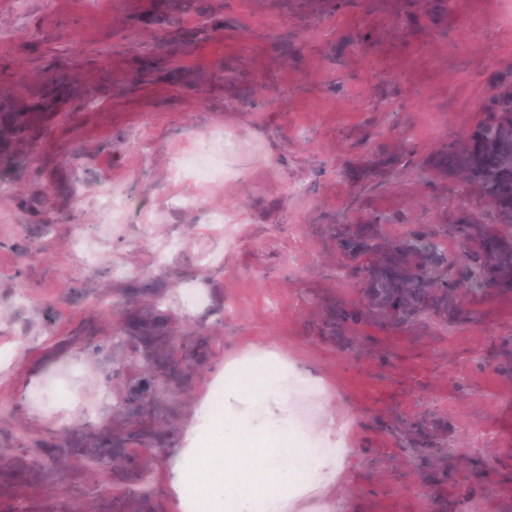}
    \vspace{-0.6em}
    \caption{Synthetic \gls{WSI}}
    \end{figure}

   \begin{figure}[h!]
    \centering
    \includegraphics[width=0.9\linewidth]{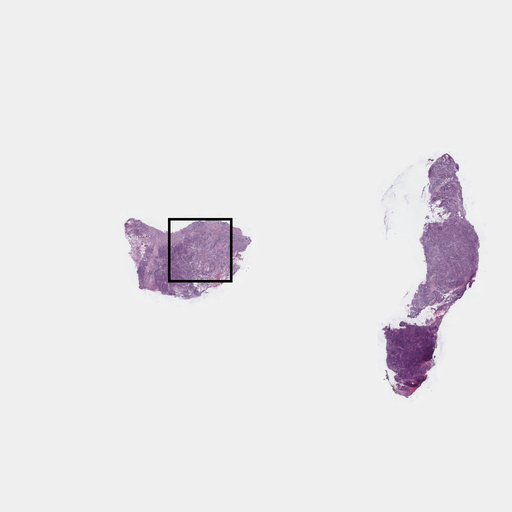}
    \vspace{0.1em}

    \includegraphics[width=0.9\linewidth]{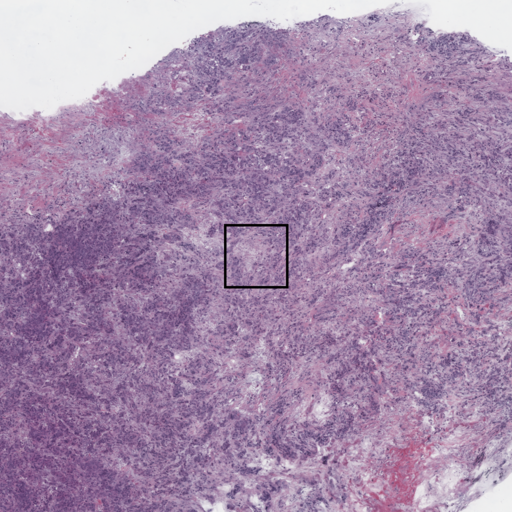}
    \vspace{0.1em}

    \includegraphics[width=0.9\linewidth]{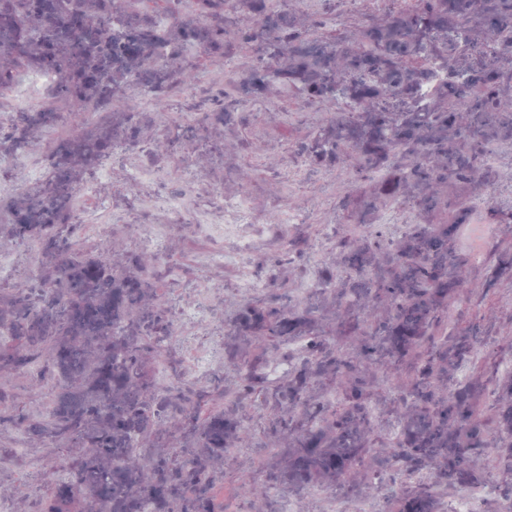}
    \vspace{-0.6em}
    \caption{Synthetic \gls{WSI}}
    \end{figure}

   \begin{figure}[h!]
    \centering
    \includegraphics[width=0.9\linewidth]{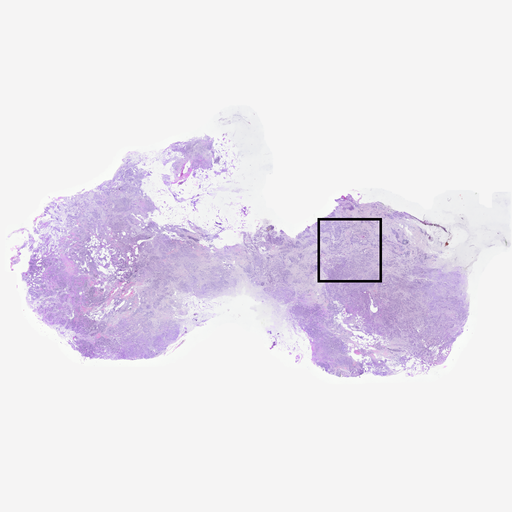}
    \vspace{0.1em}

    \includegraphics[width=0.9\linewidth]{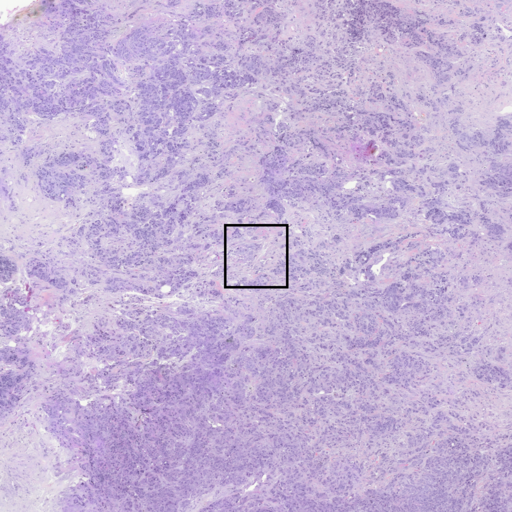}
    \vspace{0.1em}

    \includegraphics[width=0.9\linewidth]{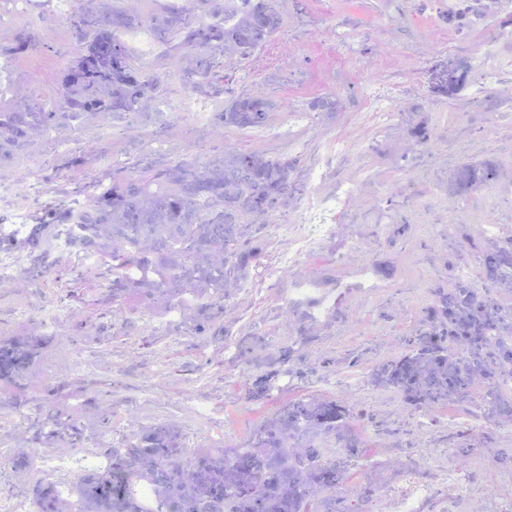}
    \vspace{-0.6em}
    \caption{Synthetic \gls{WSI}}
    \end{figure}

\end{appendix}

\end{document}